\documentclass[onecolumn,showpacs,preprintnumbers,superscriptaddress,amsmath,amssymb]{revtex4}
\usepackage[dvips]{graphicx}
\usepackage{latexsym,amssymb,amsmath,epsfig,bm,times,psfrag}
\usepackage{color}
\usepackage[bookmarksnumbered,bookmarksopen,colorlinks,citecolor=red,linkcolor=blue]{hyperref}
\usepackage{epstopdf}
\usepackage[capitalize]{cleveref}
\newcommand{\Fig}[1]{Fig.~\ref{#1}}
\newcommand{\eq}[1]{Eq.~(\ref{#1})}
\newcommand{\sect}[1]{Sec.~\ref{#1}}

\renewcommand{\part}{{\rm part}}

\newcommand{\be}{\begin{equation}}
\newcommand{\ee}{\end{equation}}
\newcommand{\bear}{\begin{eqnarray}}
\newcommand{\eear}{\end{eqnarray}}
\newcommand{\ba}{\begin{array}}
\newcommand{\ea}{\end{array}}

\begin{document}


\title{Parton evolution time dependent transverse shape asymmetry \\ in peripheral Pb-Pb and p-Pb collisions at 5.02 TeV}

\author{De-Xian Wei}
\email{dexianwei@gxust.edu.cn}
\affiliation{School of Science, Guangxi University of Science and Technology, Liuzhou, 545006, China}

\date{\today}

\begin{abstract}
In this paper, we use a multi-phase transport model to simulate the parton evolution time-dependent transverse asymmetry
in peripheral Pb-Pb (b=14-15 fm) and p-Pb (b=0-1 fm) collisions at $\sqrt{s_{NN}}$= 5.02 TeV, respectively.
The simulated results showed that the initial spatial asymmetry depends on the collision system,
while both the initial momentum asymmetry and final momentum asymmetry are independent on the collision system.
It is also shown that the final momentum asymmetry is similar to the initial momentum asymmetry.
Furthermore, the averaged transverse shape asymmetry, which includes the initial spatial asymmetry,
the initial momentum asymmetry, and the final momentum asymmetry, is significantly dependent on the transverse momentum and particle species identity (PID).
The PID scales ordering transverse asymmetry indicated that it provides a possible observation
for studying the fluctuating droplet properties of quark-gluon plasma, which may produce in heavy-ion collisions.
\end{abstract}

\pacs{25.75.Ld, 25.75.Gz}

\keywords{QGP droplets, transverse shape asymmetry, fluctuations} 
\maketitle


\section{Introduction}
\label{sec:sec1}
Collective azimuthal correlations, commonly parameterized in terms of harmonic flow coefficients $v_{n}$, were crucial experimental
signals in the discovery of the strongly interacting fluid-like quark-gluon plasma (QGP) in ultra-relativistic heavy-ion collisions.
Those collectives have been found in ultra-relativistic heavy-ion collisions
at the Relativisitic Heavy Ion Collider (RHIC) of the Brookhaven National Laboratory~\cite{Abelev:2009lrr},
and at the Large Hadron Collider (LHC) of the European Organization for Nuclear Research~\cite{ATLAS:2012mot}.

Recently, in the high multiplicity events of the small collision systems at LHC~\cite{Khachatryan:2015ltc} and RHIC~\cite{Aidala:2019coq},
long-range correlated collective signal similarly to the large collision systems have been found.
The long-range correlation experimental data of these small collision systems can be reproduced by the hydrodynamics~\cite{Mantysaari:2017iof}.
However, hydrodynamics is not the only explanation for these long-range collective correlations~\cite{Schenke:2016moo}.
The Color Glass Condensate (CGC) model can be reproduce the experimental results of these collective through gluon field correlation~\cite{Mace:2018hoa}.
Therefore, the collectives caused by small collision systems have aroused heated discussion.
These collectives observed in the small collision systems:
(1) Is it an initial state effect that one has contributed from the correlation of initial momentum in the collisions (i.e., CGC~\cite{Mace:2018hoa})?
(2) Or a final state response, that the contribution from hydrodynamic evolution~\cite{Mantysaari:2017iof}
or the contribution of other simple scattering mechanisms~\cite{Koop:2015aar,Kurkela:2019fia}?
(3) Or both of them? In experiments, people have also carried out numerous measurements,
i.e., the PHENIX of RHIC scanned energy dependent on $^{3}$He-Au and d-Au collisions,
which clearly shows that the collective behavior is a result of the final state effect~\cite{Aidala:2019coq}.
However, so far, there is still a lack of abundant and significant experimental signals to prove the complete applicability of these prediction models on small collision systems.
Although various signals tend to the final state effect, the current experiments based on small collision systems
have not yet detected other significant signals that characterize the generation of QGP droplets, i.e., jet quenching.
A suggested solution to the above-mentioned problems of QGP minimum droplets is to operate a polarized ion beam~\cite{Bozek:2018efi},
which is certainly an attractive suggestion, but it may take several years to create a suitable infrastructure.

The inherent problems of the origin of these possible QGP droplets are fascinating, and other physical properties can still be learned by studying small systems.
It is assumed that the physical mechanism behind the significance collective in Pb-Pb and p-Pb collision is the same:
they are strongly-coupled QGP substances, and satisfy the framework of thermodynamics and hydrodynamic mechanics;
the fluctuating initial geometric can successfully predict the fluctuating final collective;
there are nonequilibrium partial sub-dynamics (it may cause the system to evolve from a weak-coupled collective to a strong-coupled collective).
The hydrodynamics points out that~\cite{Noronha-Hostler:2016sof} the influence of subnucleon color fluctuations in Pb-Pb and p-Pb collisions is different,
indeed, eccentricity in p-Pb collision is sensitive to subnucleon scale fluctuations,
and the collective in Pb-Pb collision with the same multiplicity is not particularly sensitive to microscopic subnucleon physics.
Therefore, subnucleon fluctuations are critical to understanding the collective correlations in small collision systems.
Recently, it has been recognized that hot spot fluctuations in subnucleons are crucial to understanding the geometry of the initial stage of high energy collisions~\cite{Mantysaari:2020rop}.
It leads to a series of new studies on the transverse spatial distribution of freedom degrees of subnucleon in protons and nuclei~\cite{Schlichting:2014tso,Mantysaari:2016eos}.
Ref.~\cite{Demirci:2021hsa} reveals the main source of fluctuation on the subnucleon scale:
the geometric fluctuation of a hot spot in proton (projectile side) is the main source of eccentricity,
while the fluctuation of color (target side) can only give a negligible correction.
Follow this subject, we study the asymmetry degrees in symmetrical small collisions and asymmetrical small collisions in this paper.

While, there are still not clear enough about geometric fluctuations of time-dependent evolution, especially in the process of nonequilibrium dynamics.
If we assume that a collective exists in small collision systems, the fluctuating behavior of these collective represent
critical tool for studying the properties of the QGP medium.
To explore these fluctuations, in this paper,
we mainly start from the time-dependent transverse asymmetry to study the general geometric fluctuation properties of the peripheral Pb-Pb and p-Pb collision systems.
To this purpose, we focus on time-dependent transverse shape asymmetry in peripheral Pb-Pb and p-Pb collisions using an
event averaged analysis.
This paper does not intend to talk about collective flow and non-collective behavior in small collision systems.
Notice that in this paper, we do not attempt to compare simulations with experimental data,
but rather to explore how the transverse asymmetry is influenced by the dynamic and/or dynamic fluctuations.

The paper is organized as follows: In \sect{sec:sec2} we briefly describe
the events averaged x-y and transverse components of asymmetry in a multi-phase transport (AMPT)
model~\cite{Lin:2004amt}, which is then used in the simulations.
Numerical results about the events averaged x-y and transverse components of asymmetry are presented in \sect{sec:sec3}.
Finally, we summarize the main results in \sect{sec:sum}.

\section{Materials and Methods}
\label{sec:sec2}

One observation to probe the transverse subnucleon fluctuations is the transverse shape asymmetry, $A(x)$.
In a single event, the transverse shape asymmetry is defined as
\begin{eqnarray}\label{asym:def1}
A(x) &=& \frac{\sqrt{\sum_{i=1}^{N}(x_{i}-\langle x\rangle)^{2}}}{\sqrt{N}\langle x\rangle}.
\end{eqnarray}
where brackets $\langle \cdots \rangle$ indicate the present particles are averaged over the entire event and $N$ is the multiplicity in a single event.
The variables $x$ can be considered samples from the spatial or the momentum in evolutionary stage.
In this case, $A(x)=0$ means that
the samples are completely symmetrical, whereas and $A(x)\neq 0$ corresponds to asymmetrical quantities.
For the event-by-event simulations, the events averaged transverse asymmetry is expressed as
\begin{eqnarray}\label{asym:def2}
\langle A(x)\rangle &=& \langle \frac{\sqrt{\sum_{i=1}^{N}(x_{i}-\langle x\rangle)^{2}}}{\sqrt{N}\langle x\rangle}\rangle.
\end{eqnarray}

To estimate the whole particle of an averaged method, one must be careful about noncollective physics in the strong color-coupled system,
i.e., non-hydrodynamic-like/particle-like, which contributed from the jet fragment and hadron resonance.
While, in this paper, we focus only on the fluctuation evolved with time in the strong couple system, not about the collective of the system.
Through this paper, we consider all charged particles in the present range (with 0.3 $< p_{T} <$ 3.0 GeV and $|\eta|<$2.40).

In this work, the transverse shape asymmetry is investigated in peripheral Pb-Pb (b=14-15 fm) and p-Pb (b=0-1 fm) collisions
at $\sqrt{s_{NN}}$ = 5.02 TeV by the AMPT model~\cite{Lin:2004amt}, respectively.
Note that the impact parameters are controlled by the transverse distances of the overlap in initial collision space,
not the final charged participant multiplicity, $M$. As a consequence, it includes a wide multiplicity range, from low $M$ to large ones.
This paper takes the specific shear viscosity $\eta/s=0.273$,
one is calculated by the Lund string fragmentation parameters in AMPT~\cite{De:2020rli}, i.e., $a=0.5$, $b=0.9$ GeV$^{-2}$,
$\alpha_{s}$ =0.33 and $\mu$ = 3.2 fm$^{-1}$. More thorough details of the AMPT model can be found in Ref.~\cite{Lin:2004amt}.
The results for Pb-Pb and p-Pb collisions are obtained by simulating 5$\times 10^{5}$ events for each given impact parameter.

\section{Results}
\label{sec:sec3}

\begin{figure*}[tp]
\begin{center}
\includegraphics[width=0.320\textwidth]{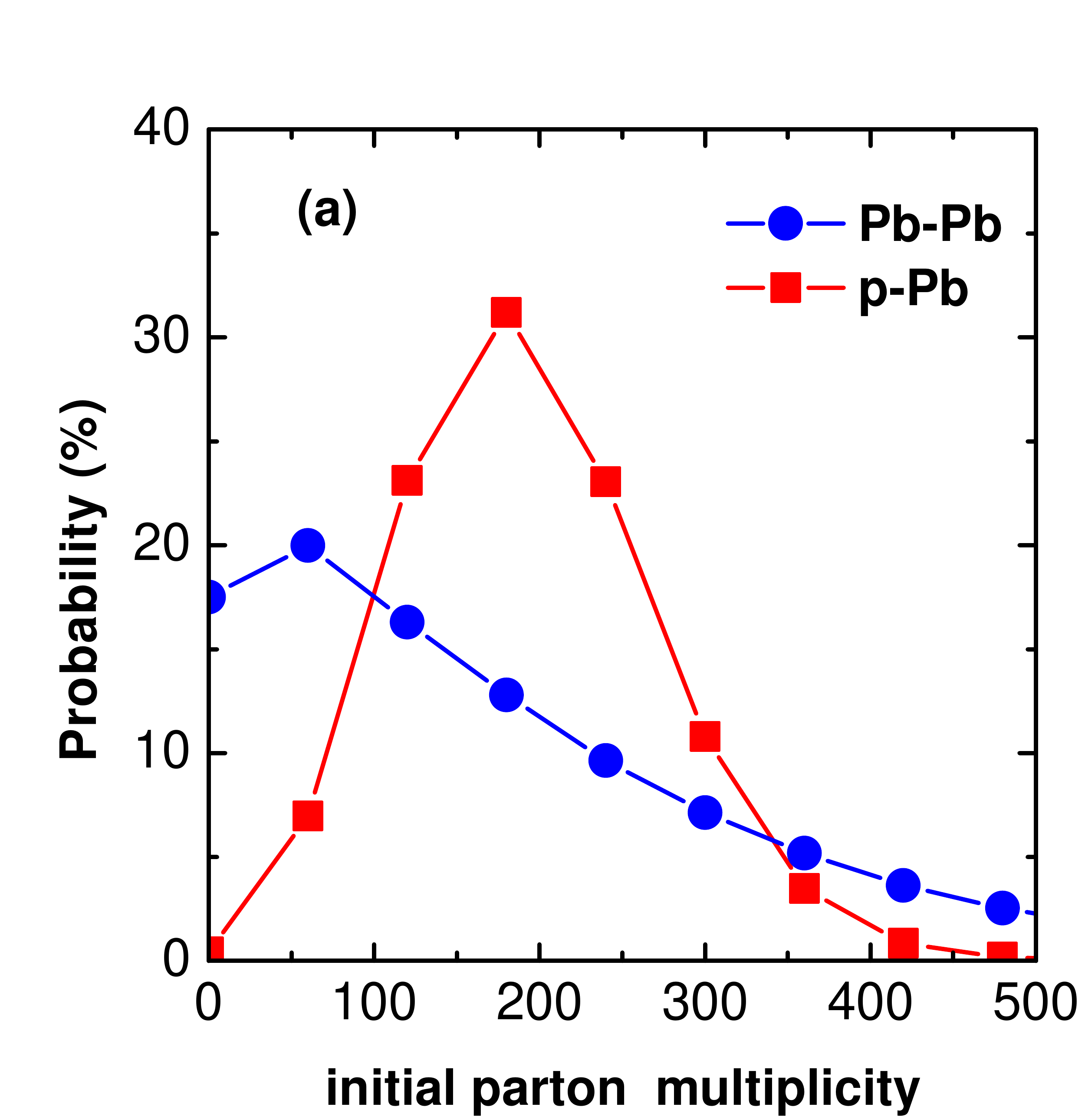}
\hspace{0.80cm}
\includegraphics[width=0.330\textwidth]{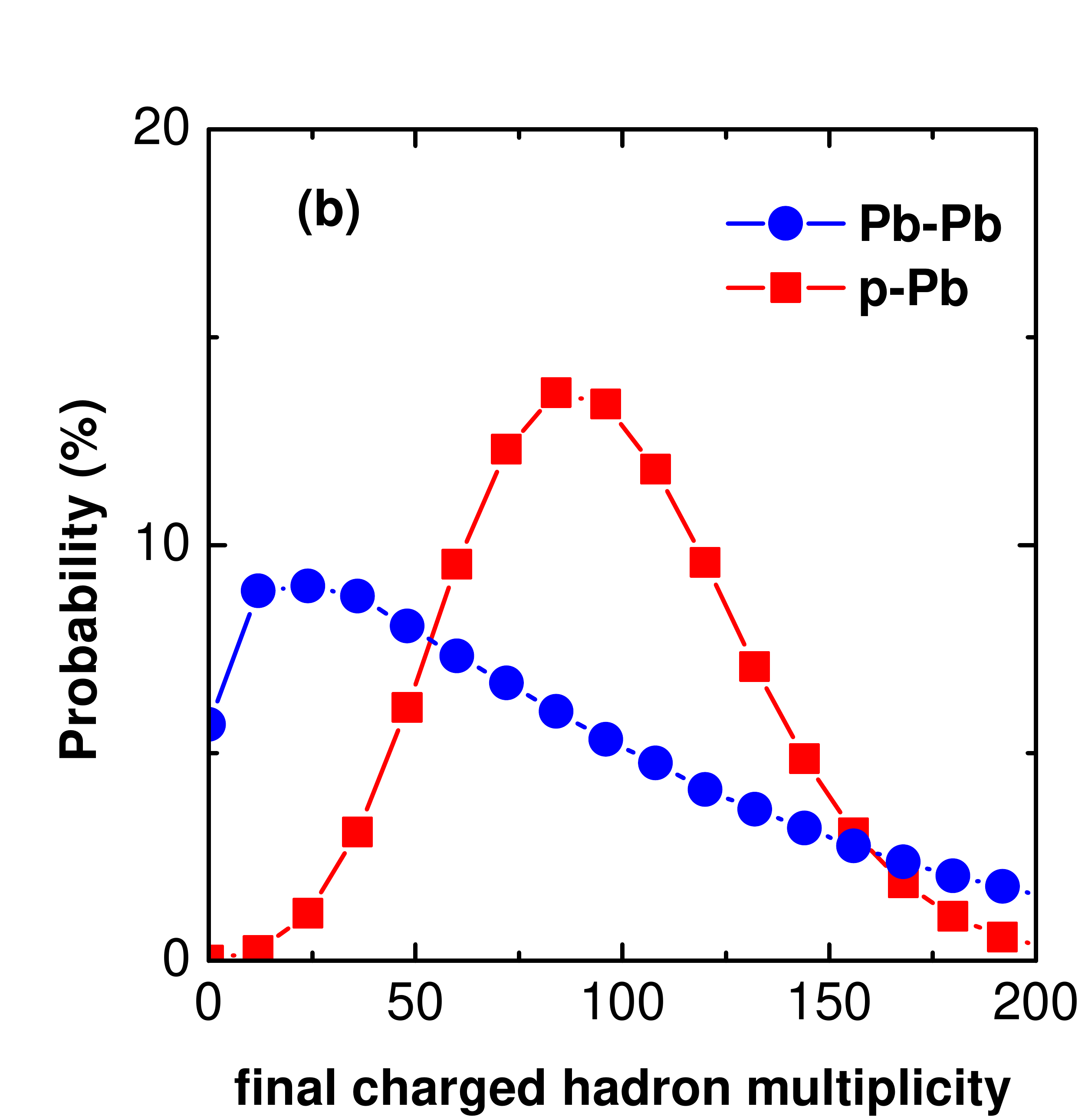}\\
\includegraphics[width=0.320\textwidth]{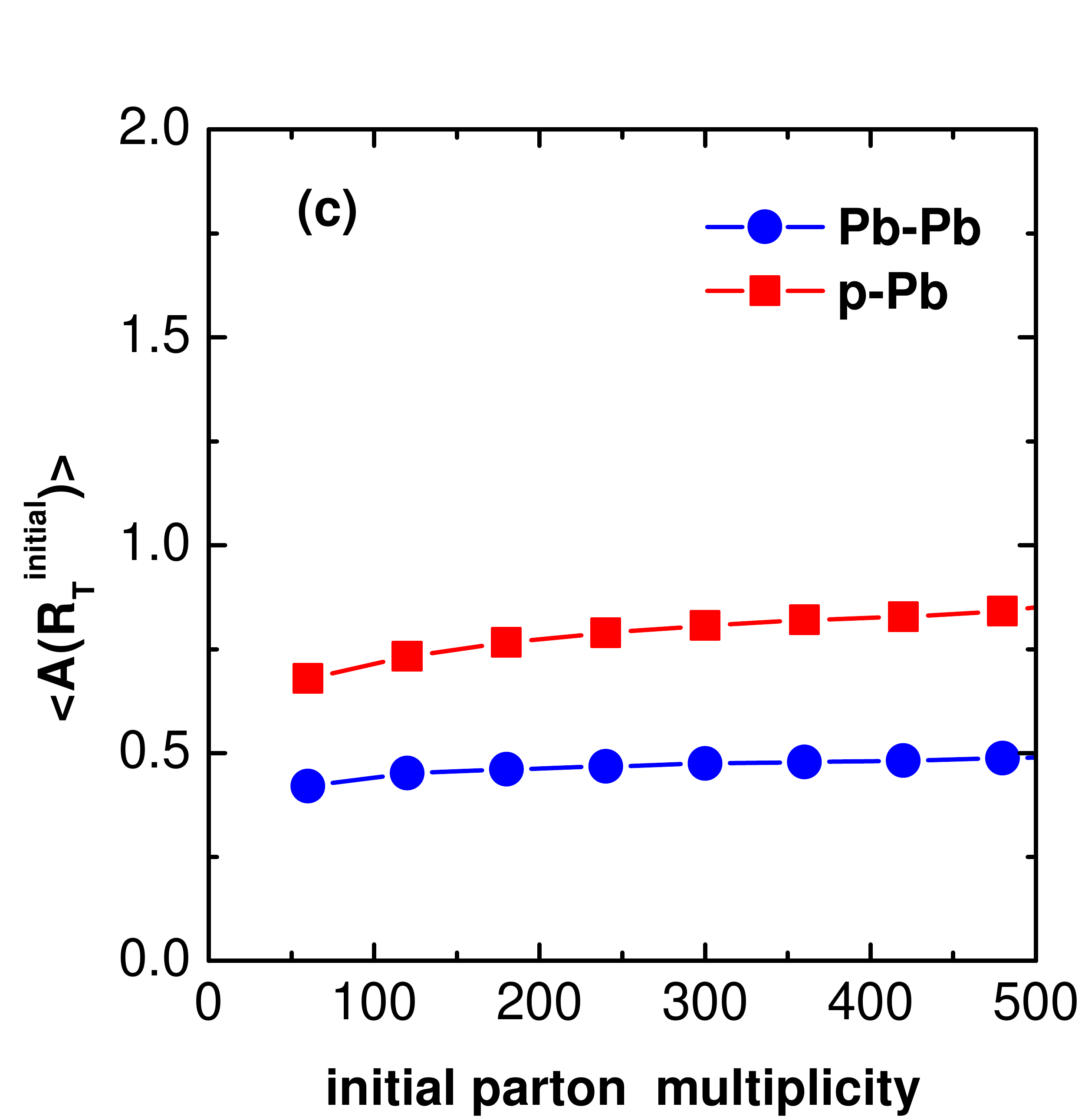}
\hspace{0.80cm}
\includegraphics[width=0.330\textwidth]{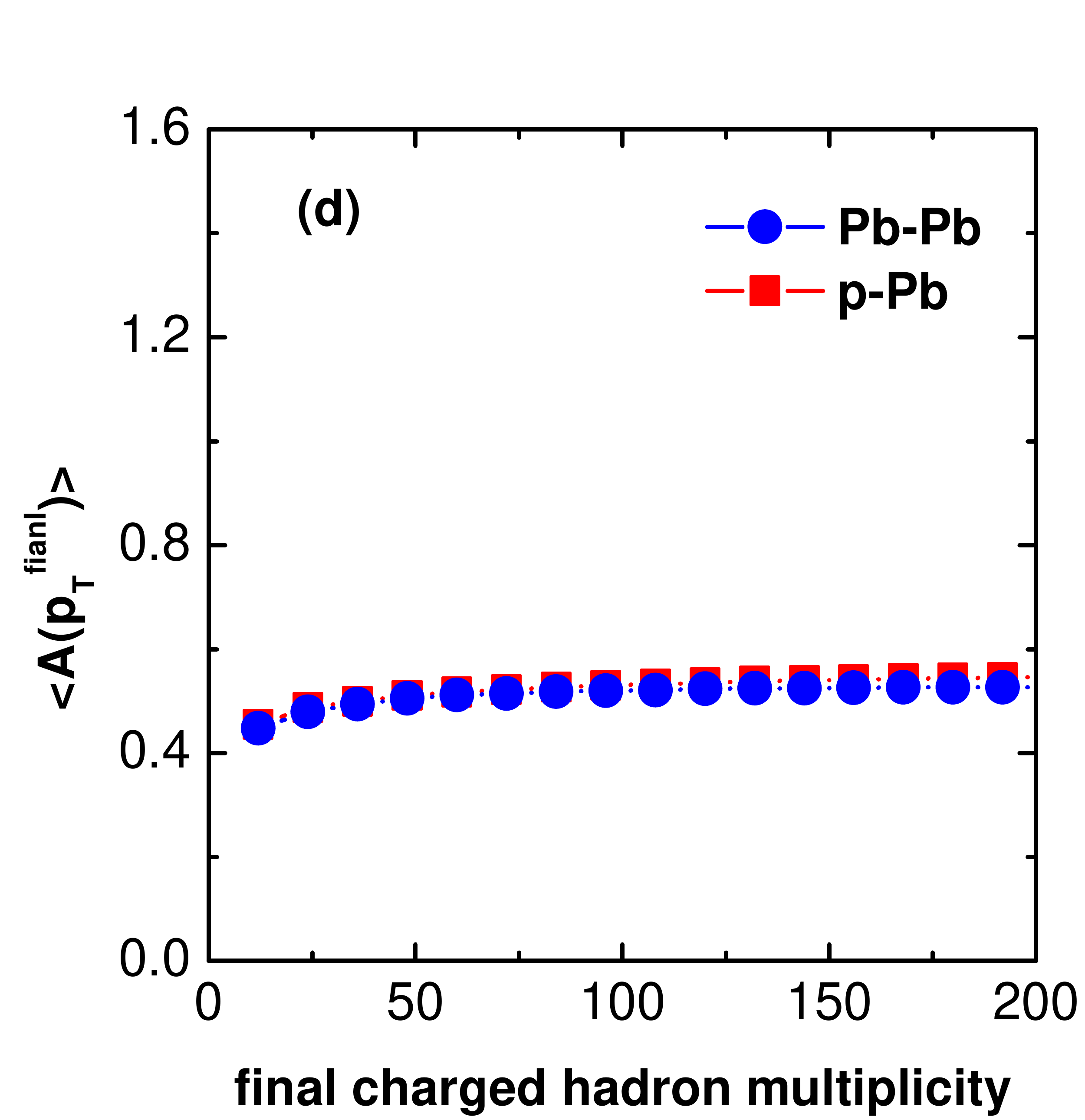}
\caption{(Color online)
Up panels: (a) the probability of initial parton multiplicity and (b) the probability of final charged hadron multiplicity
in Pb-Pb (b=14-15 fm) and p-Pb (b=0-1 fm) collisions at $\sqrt{s_{NN}}$= 5.02 TeV, respectively.
Down panels: (c) the initial averaged transverse spatial asymmetry as a function of the initial parton multiplicity and
(d) the final averaged transverse momentum asymmetry as a function of the final charged hadron multiplicity,
in Pb-Pb (b=14-15 fm) and p-Pb (b=0-1 fm) collisions at $\sqrt{s_{NN}}$= 5.02 TeV, respectively.
}
\label{fig1}
\end{center}
\end{figure*}

\begin{figure*}[h]
\begin{center}
\includegraphics[width=0.30\textwidth]{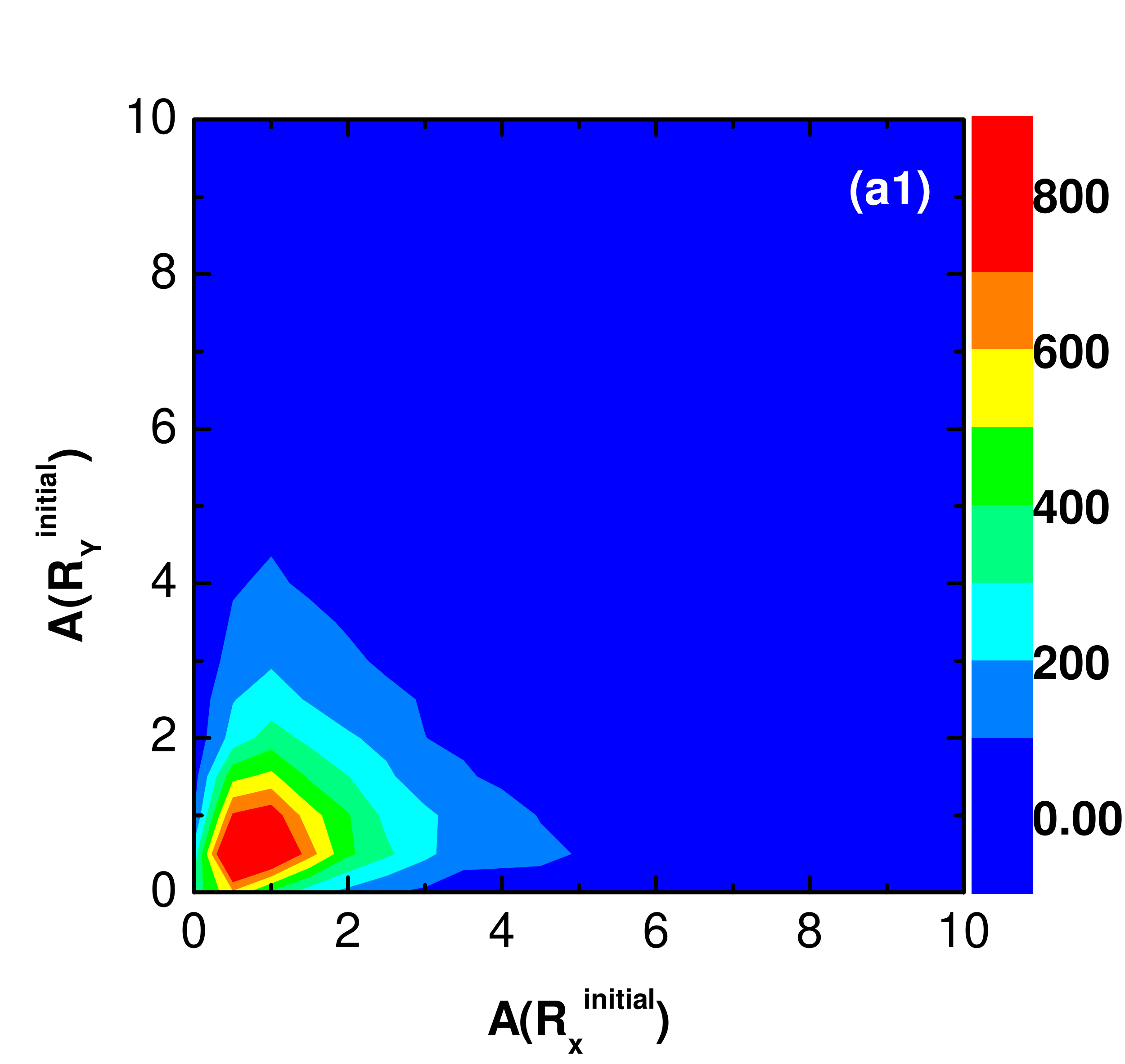}
\hspace{0.25cm}
\includegraphics[width=0.30\textwidth]{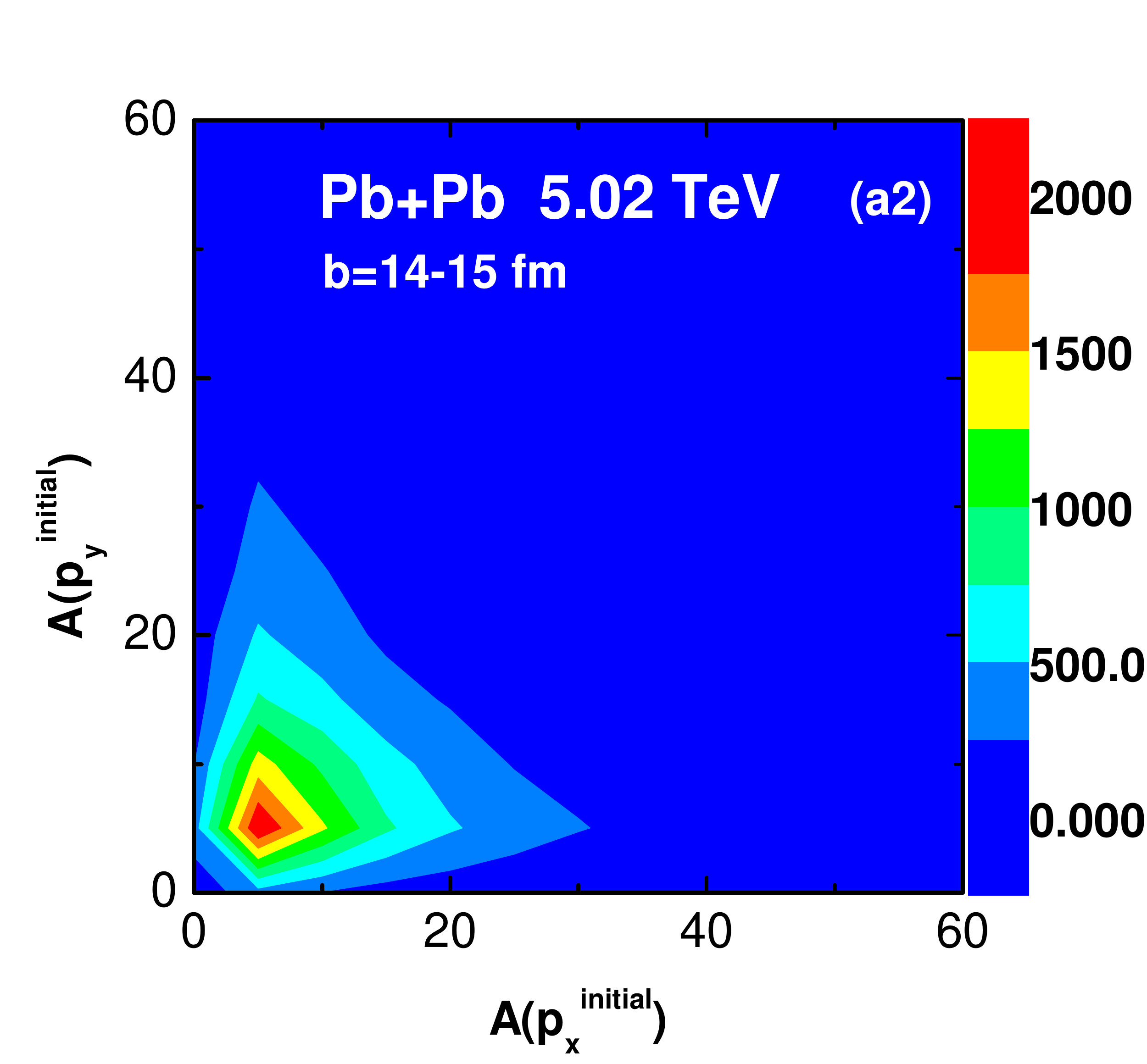}
\hspace{0.25cm}
\includegraphics[width=0.30\textwidth]{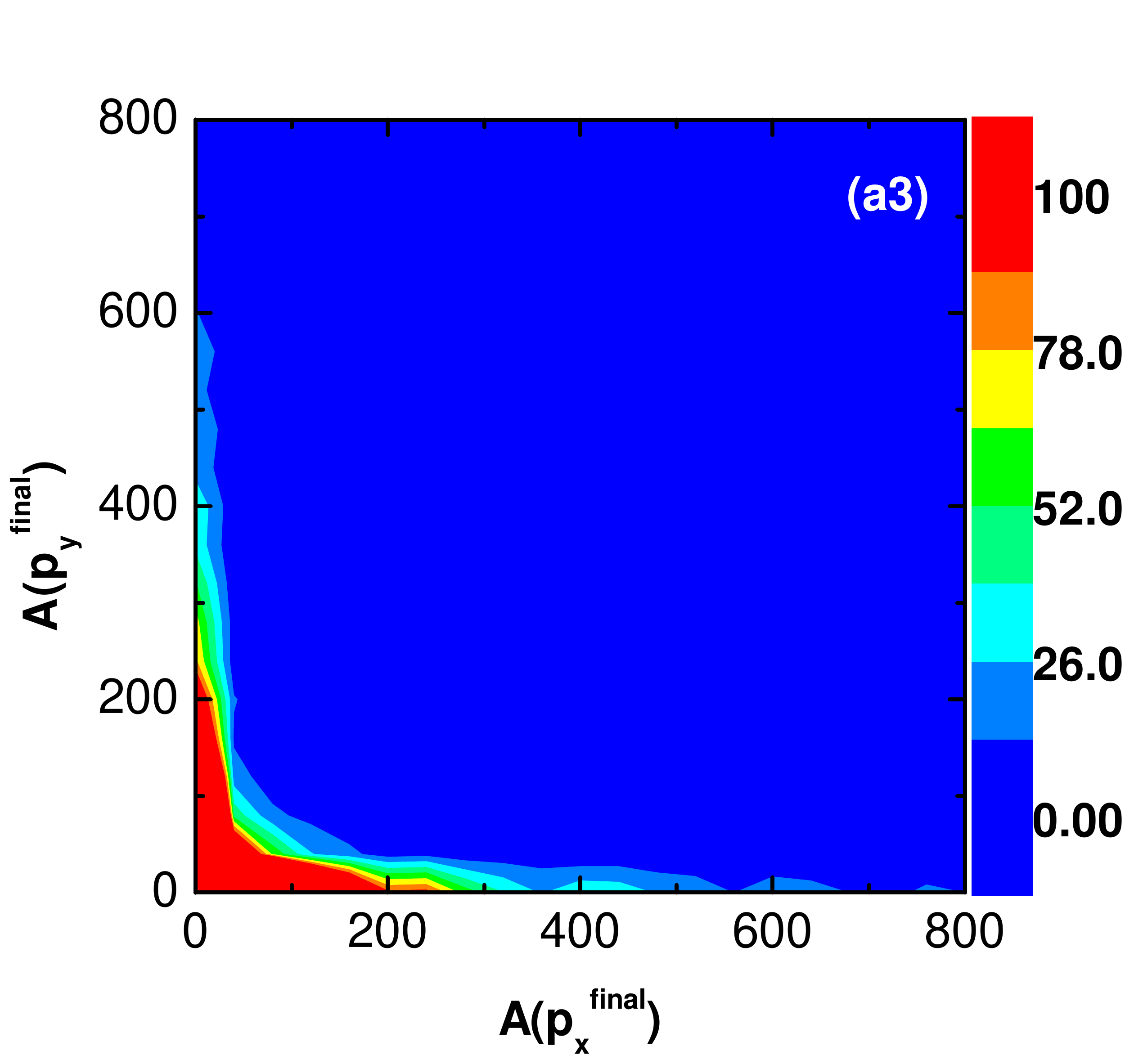}\\
\includegraphics[width=0.30\textwidth]{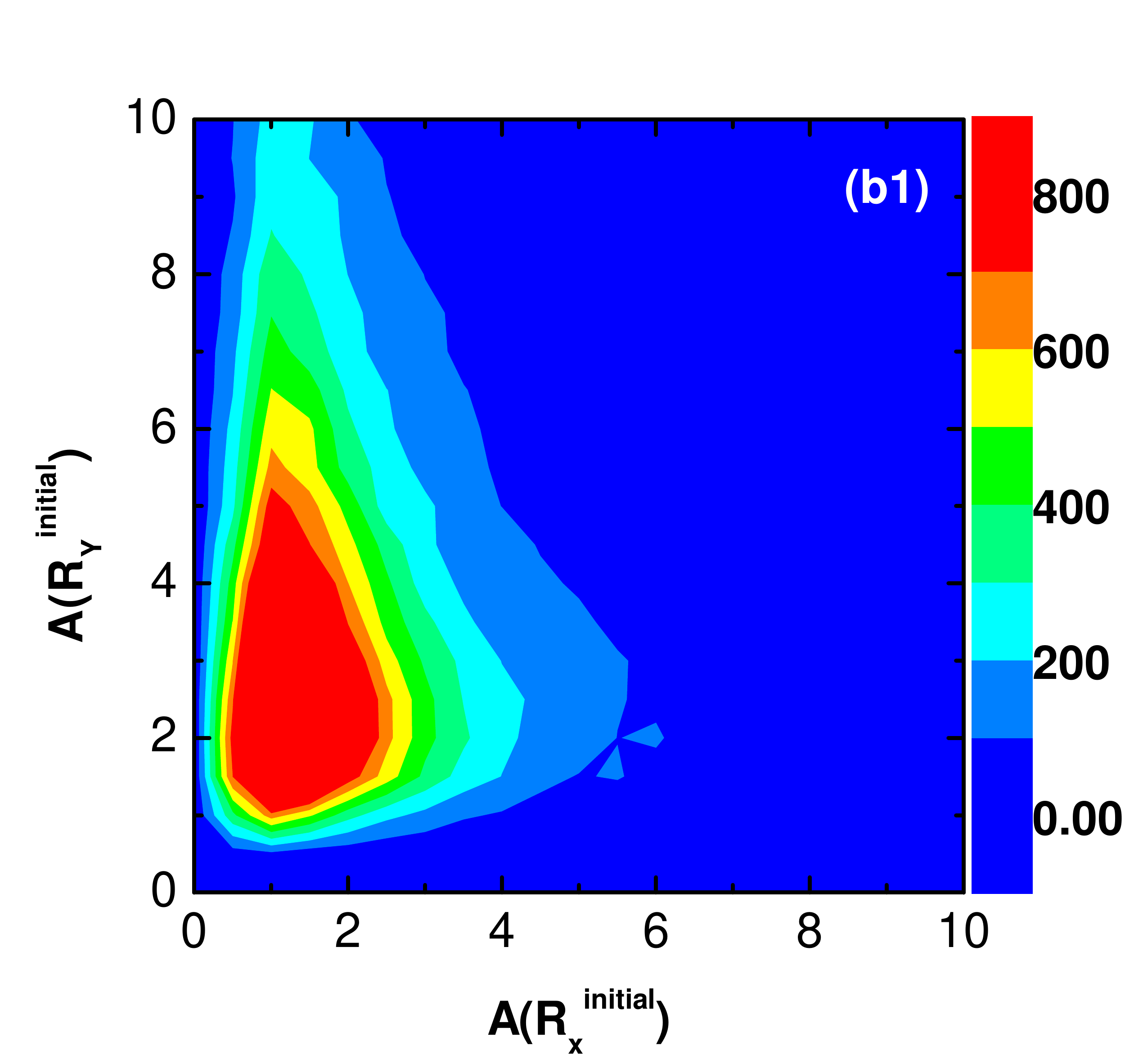}
\hspace{0.25cm}
\includegraphics[width=0.30\textwidth]{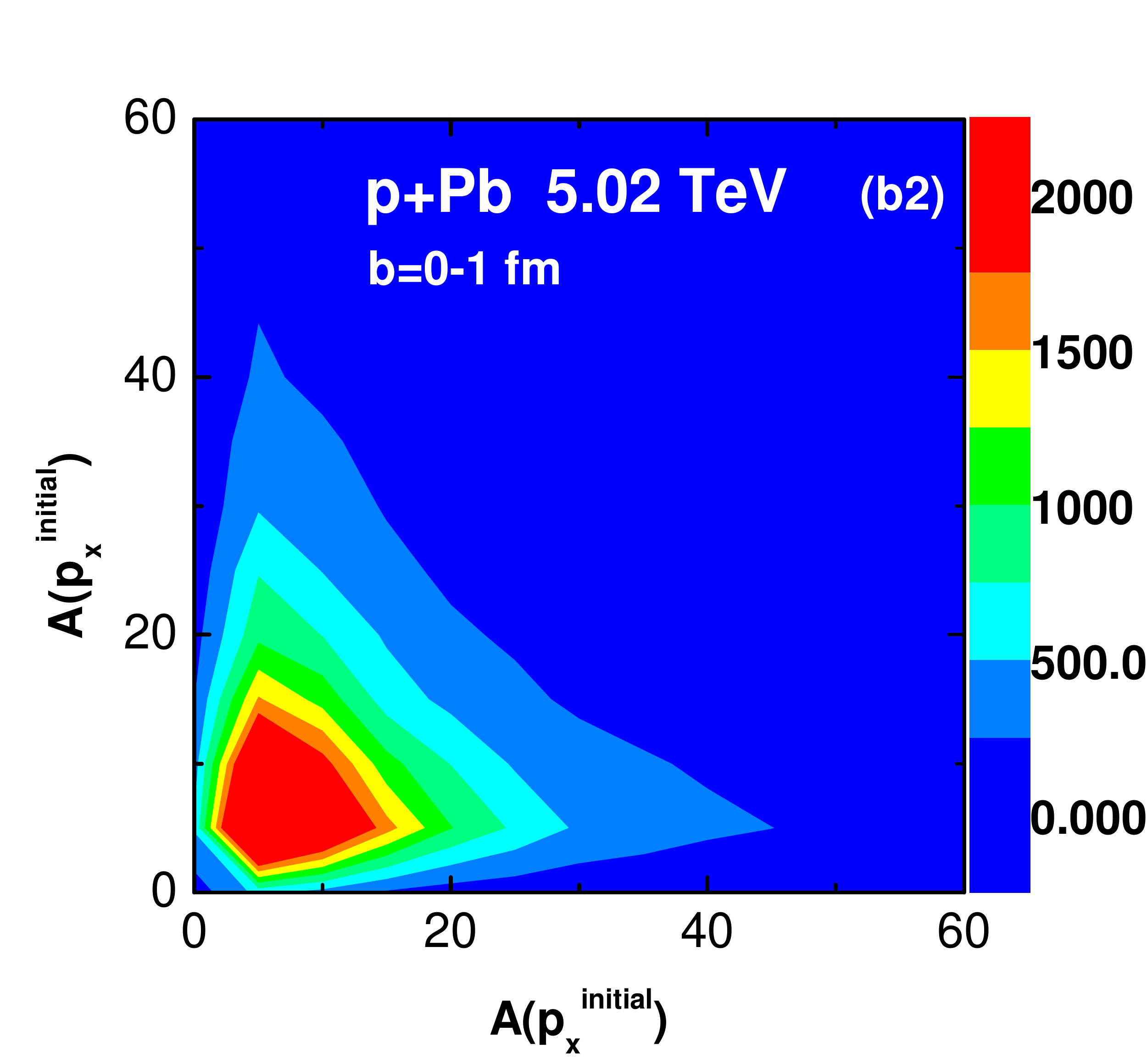}
\hspace{0.25cm}
\includegraphics[width=0.30\textwidth]{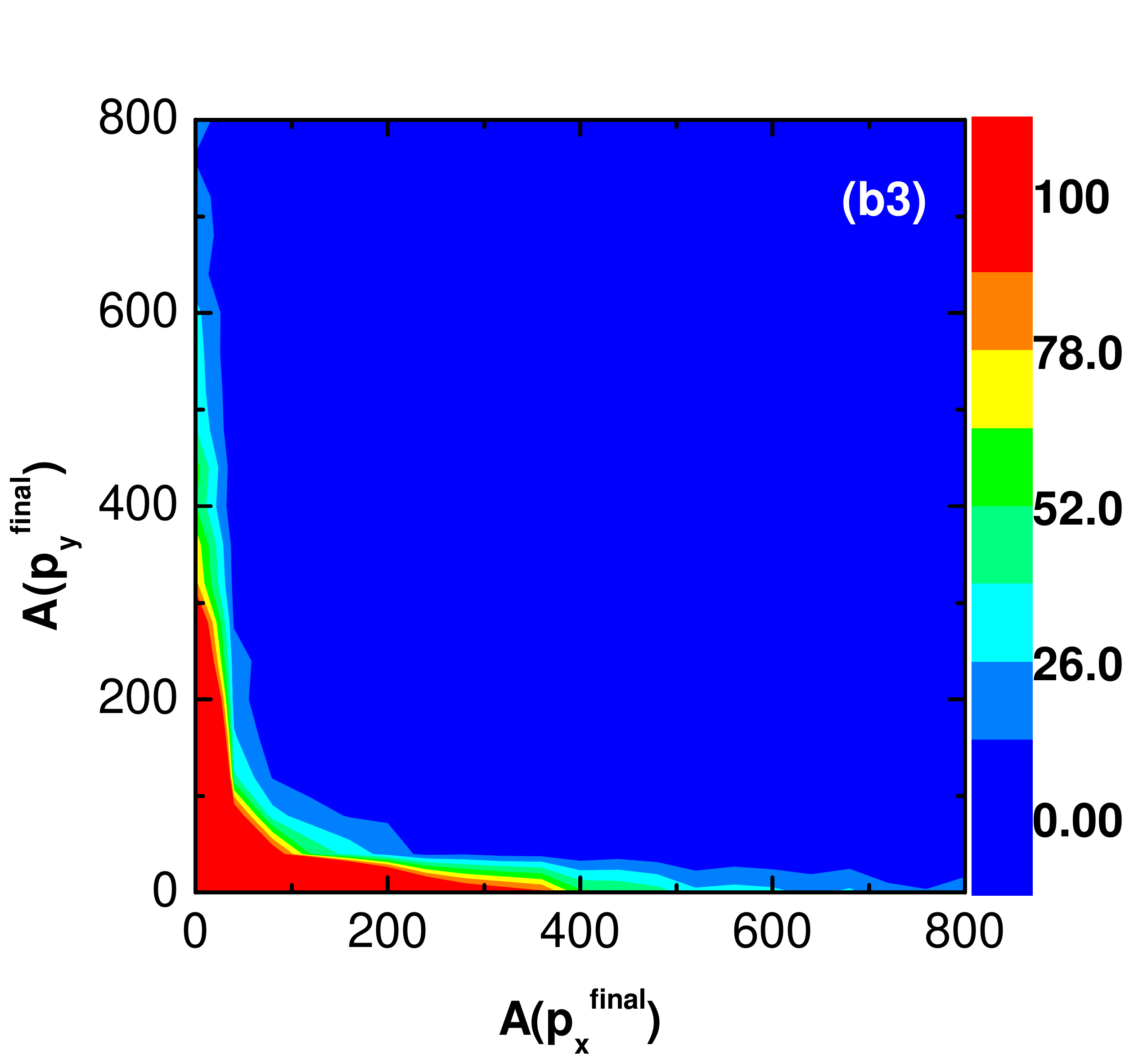}
\caption{(Color online)
The x-y components of the asymmetry distributions of initial spatial, initial momentum, and final momentum are shown in Pb-Pb (b=14-15 fm) and p-Pb (b=0-1 fm) collisions, respectively.
Up panels: for Pb-Pb collisions. Down panels: for p-Pb collisions.
}
\label{fig2}
\end{center}
\end{figure*}

\begin{figure*}[h]
\begin{center}
\includegraphics[width=0.36\textwidth]{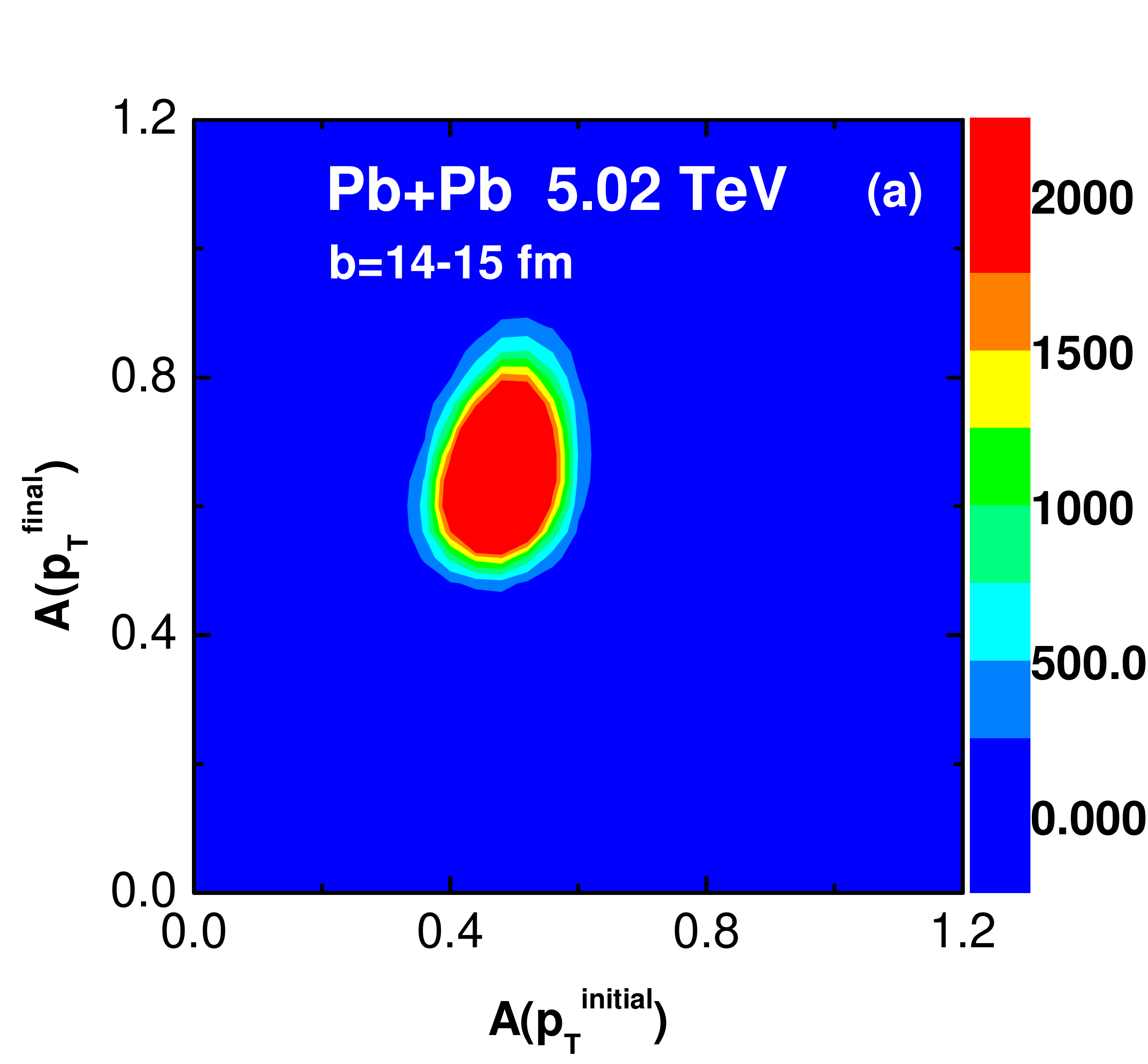}
\hspace{1.00cm}
\includegraphics[width=0.36\textwidth]{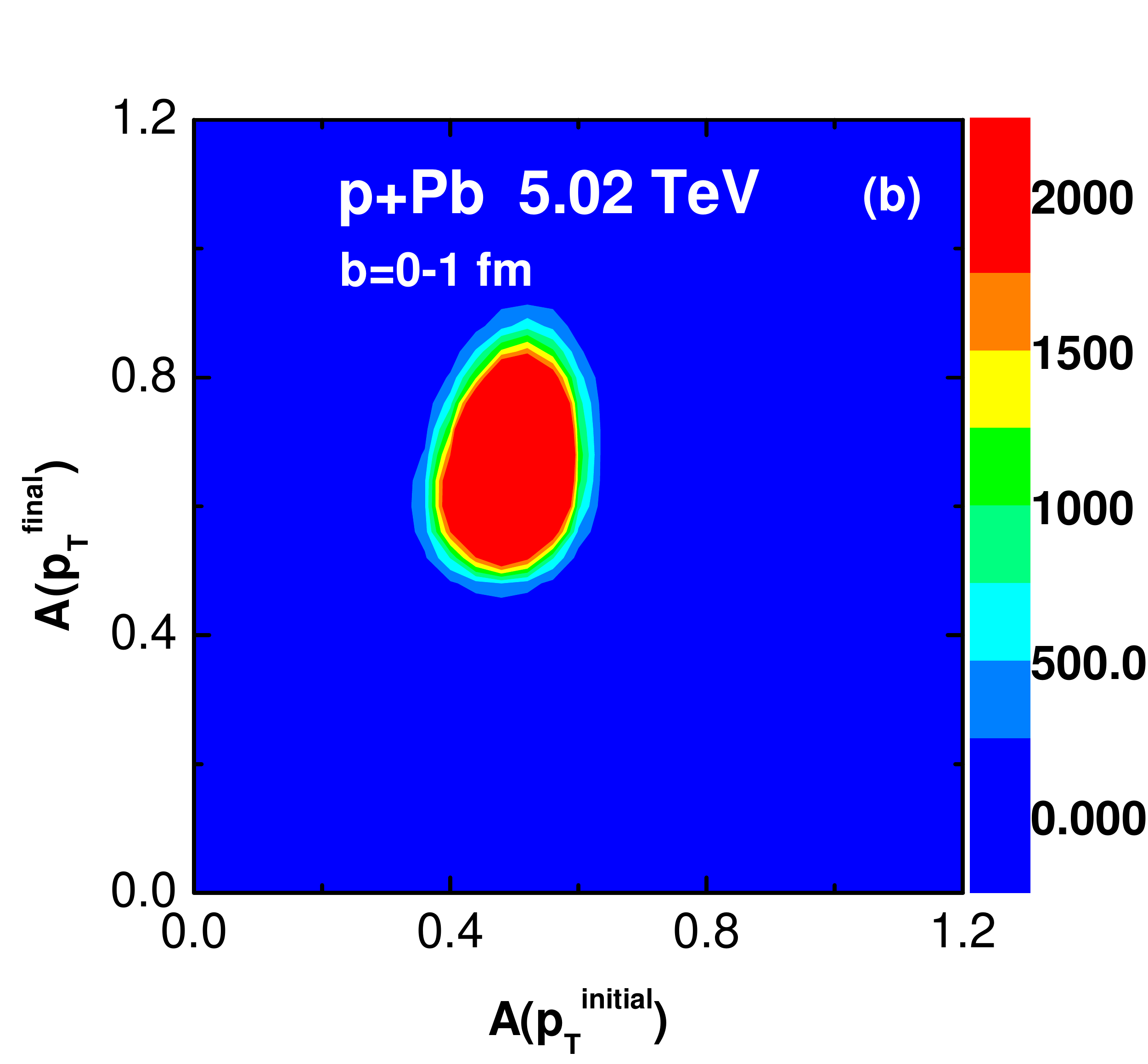}
\caption{(Color online)
The transverse asymmetry distributions between final momentum and initial momentum are shown in Pb-Pb (b=14-15 fm) and p-Pb (b=0-1 fm) collisions, respectively.
Left panel: for Pb-Pb collisions. Right panel: for p-Pb collisions.
}
\label{fig3}
\end{center}
\end{figure*}

\begin{figure*}[h]
\begin{center}
\includegraphics[width=0.30\textwidth]{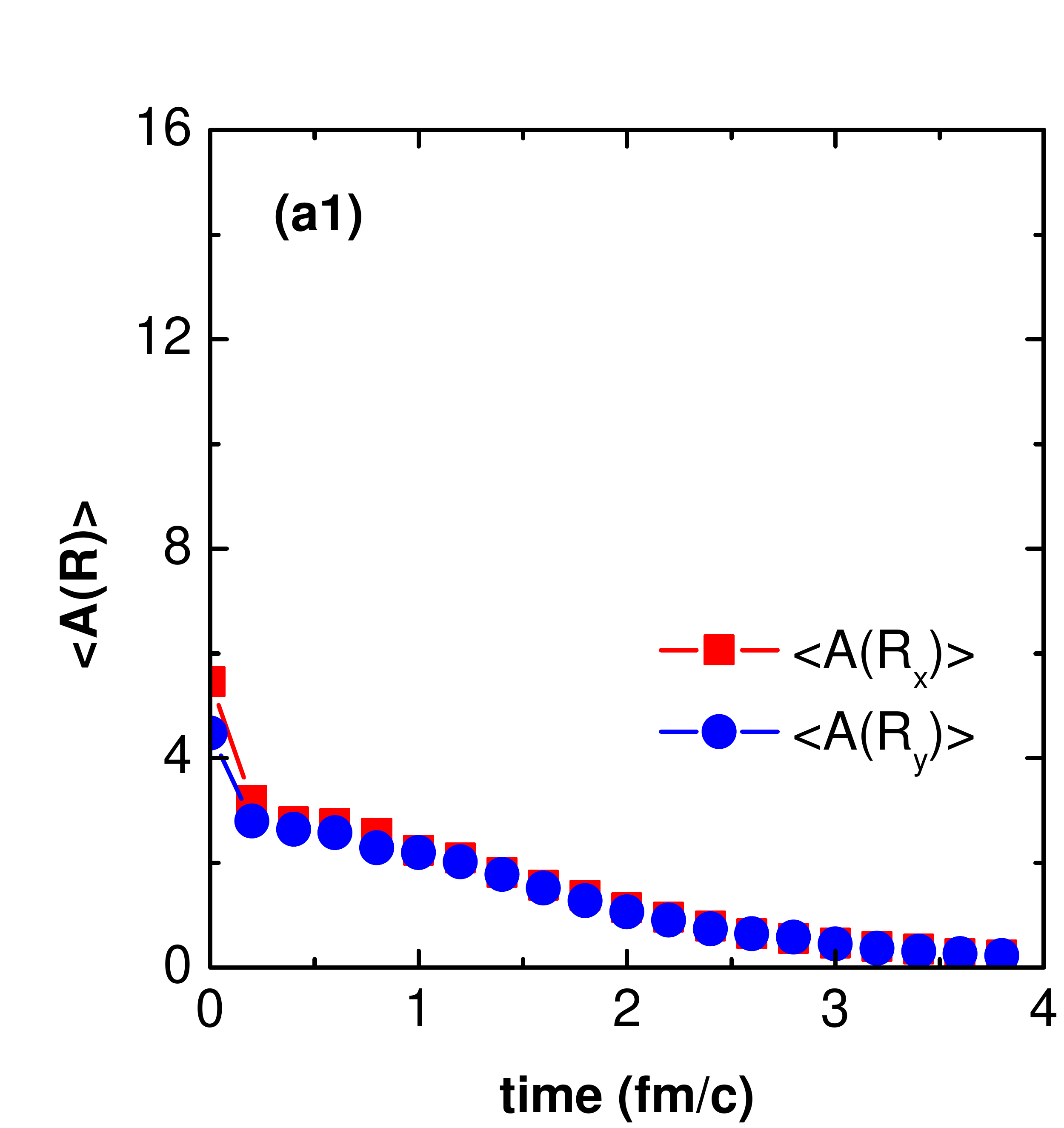}
\hspace{0.25cm}
\includegraphics[width=0.30\textwidth]{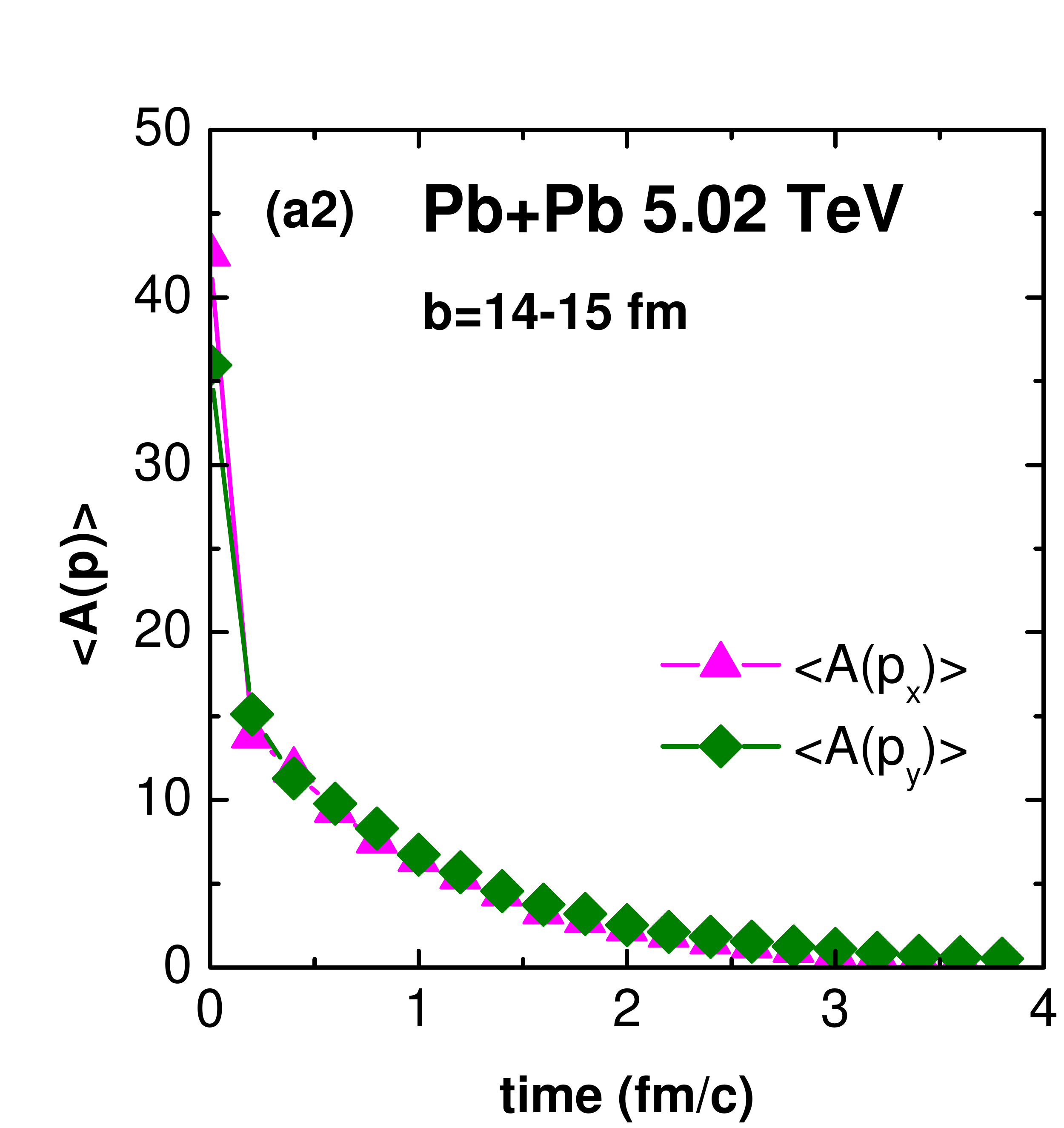}
\hspace{0.25cm}
\includegraphics[width=0.30\textwidth]{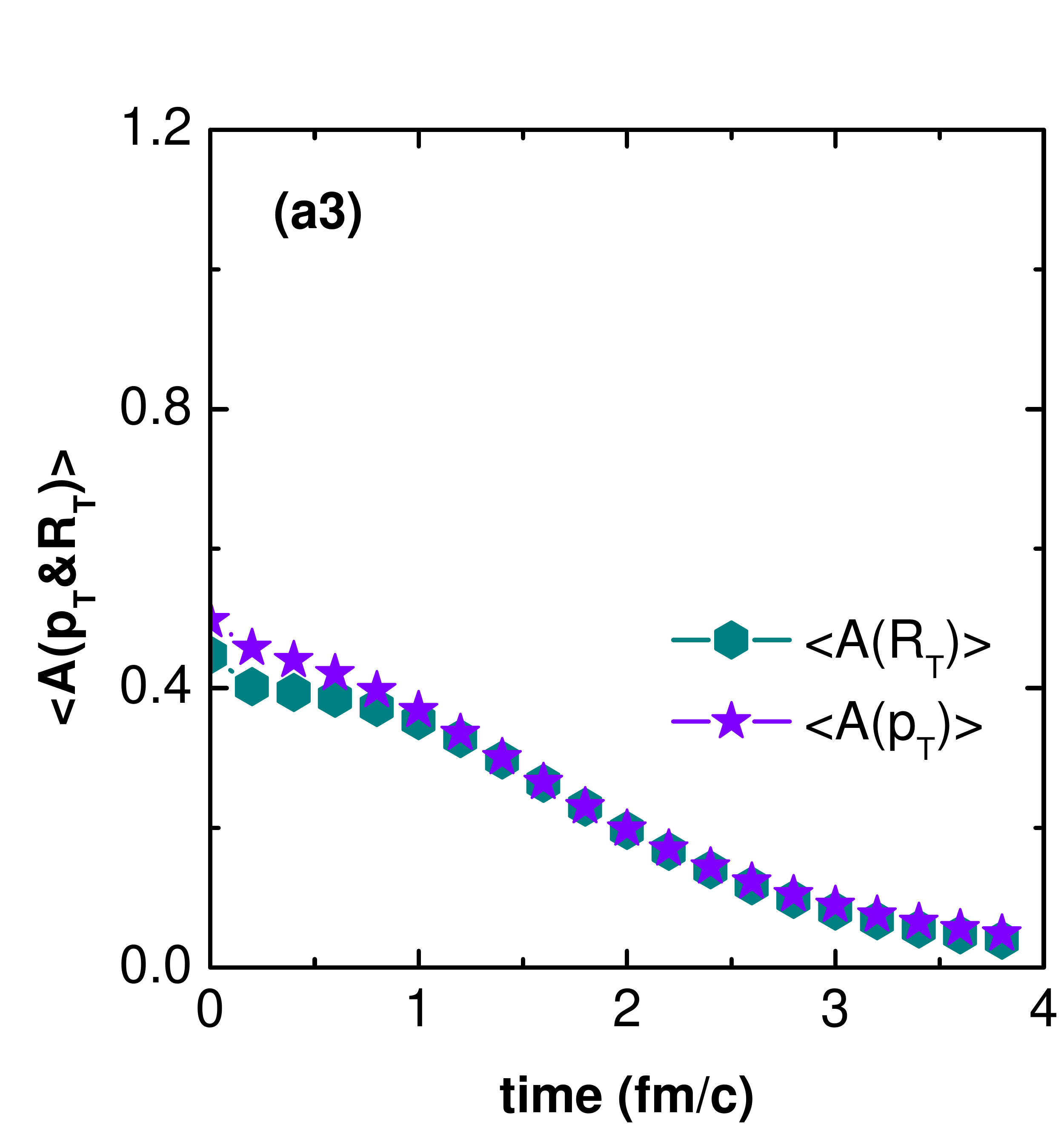}\\
\includegraphics[width=0.30\textwidth]{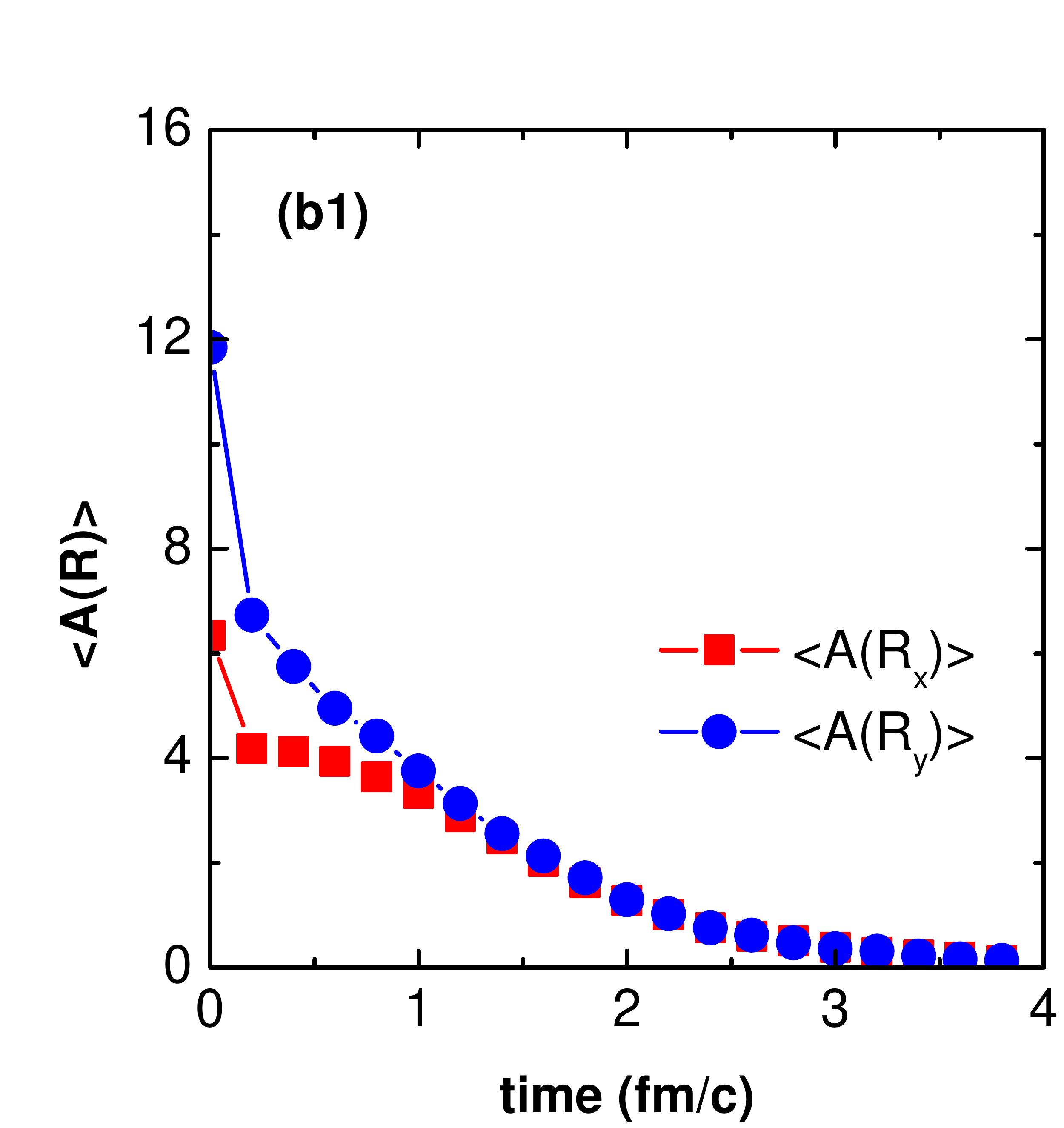}
\hspace{0.25cm}
\includegraphics[width=0.30\textwidth]{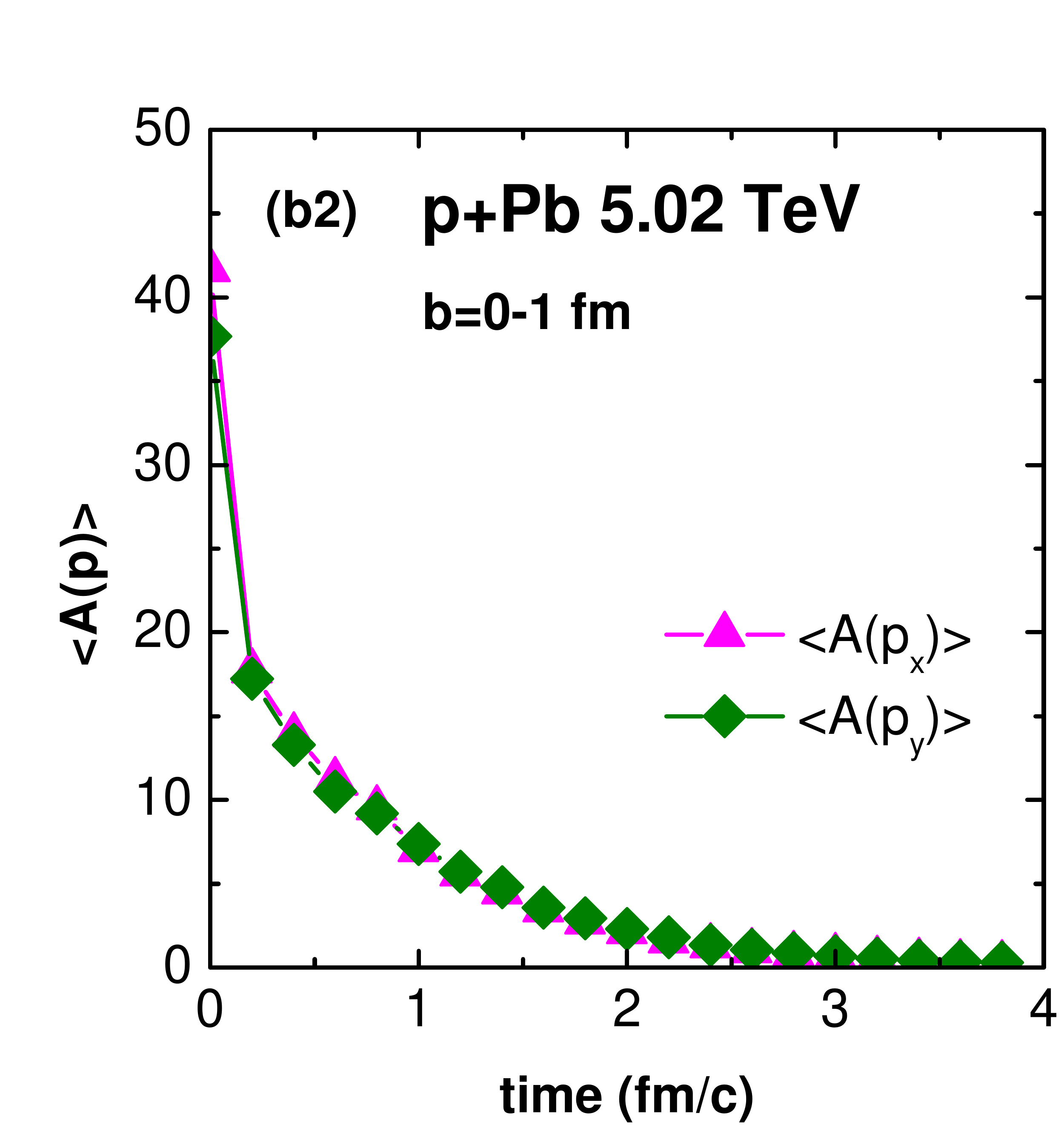}
\hspace{0.25cm}
\includegraphics[width=0.30\textwidth]{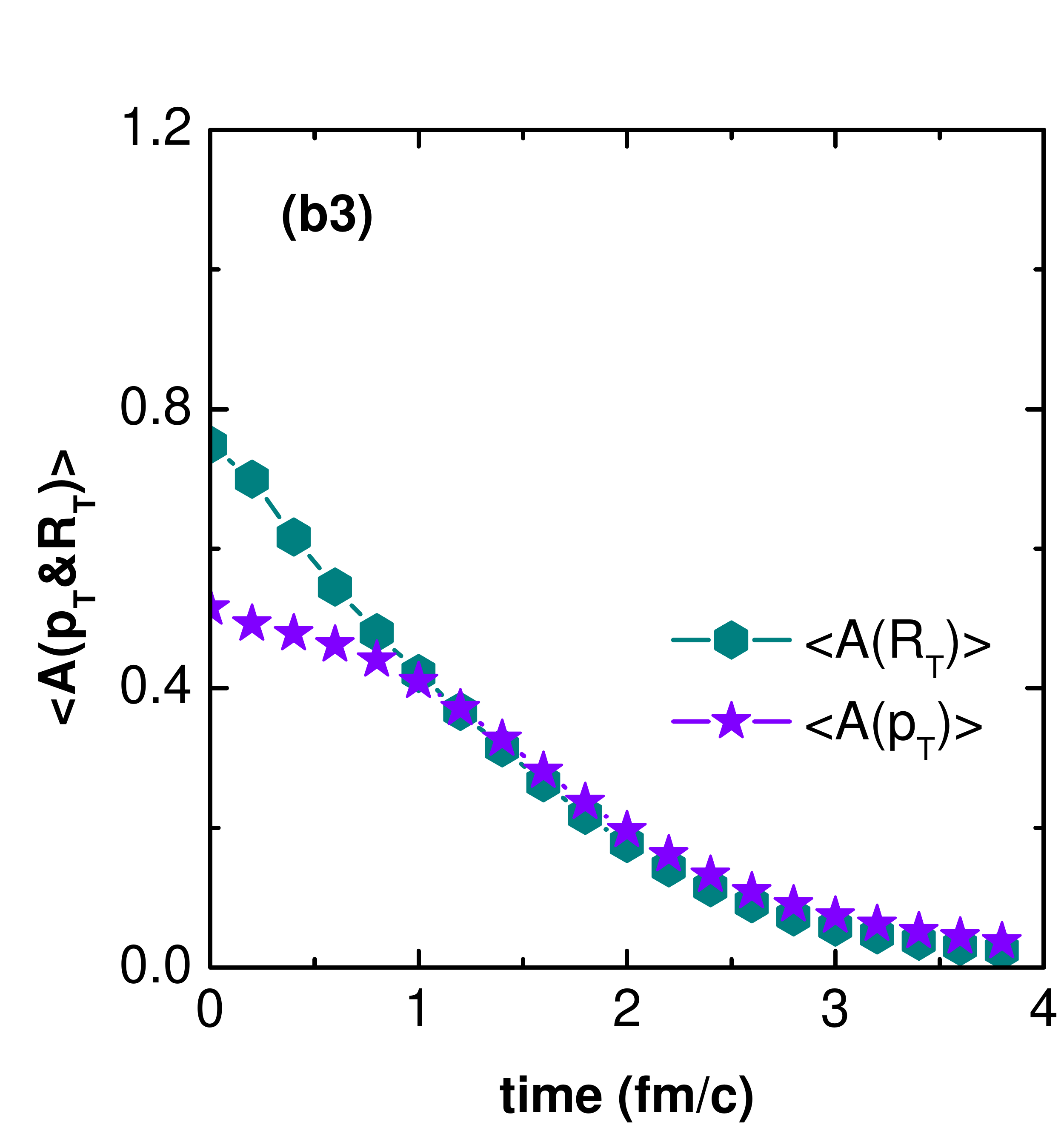}
\caption{(Color online)
The x-y and transverse components of averaged asymmetry of spatial and momentum as functions of the parton evolution time.
Up panels: for Pb-Pb (b=14-15 fm) collisions. Down panels: for p-Pb (b=0-1 fm) collisions.
}
\label{fig4}
\end{center}
\end{figure*}

\begin{figure*}[h]
\begin{center}
\includegraphics[width=0.360\textwidth]{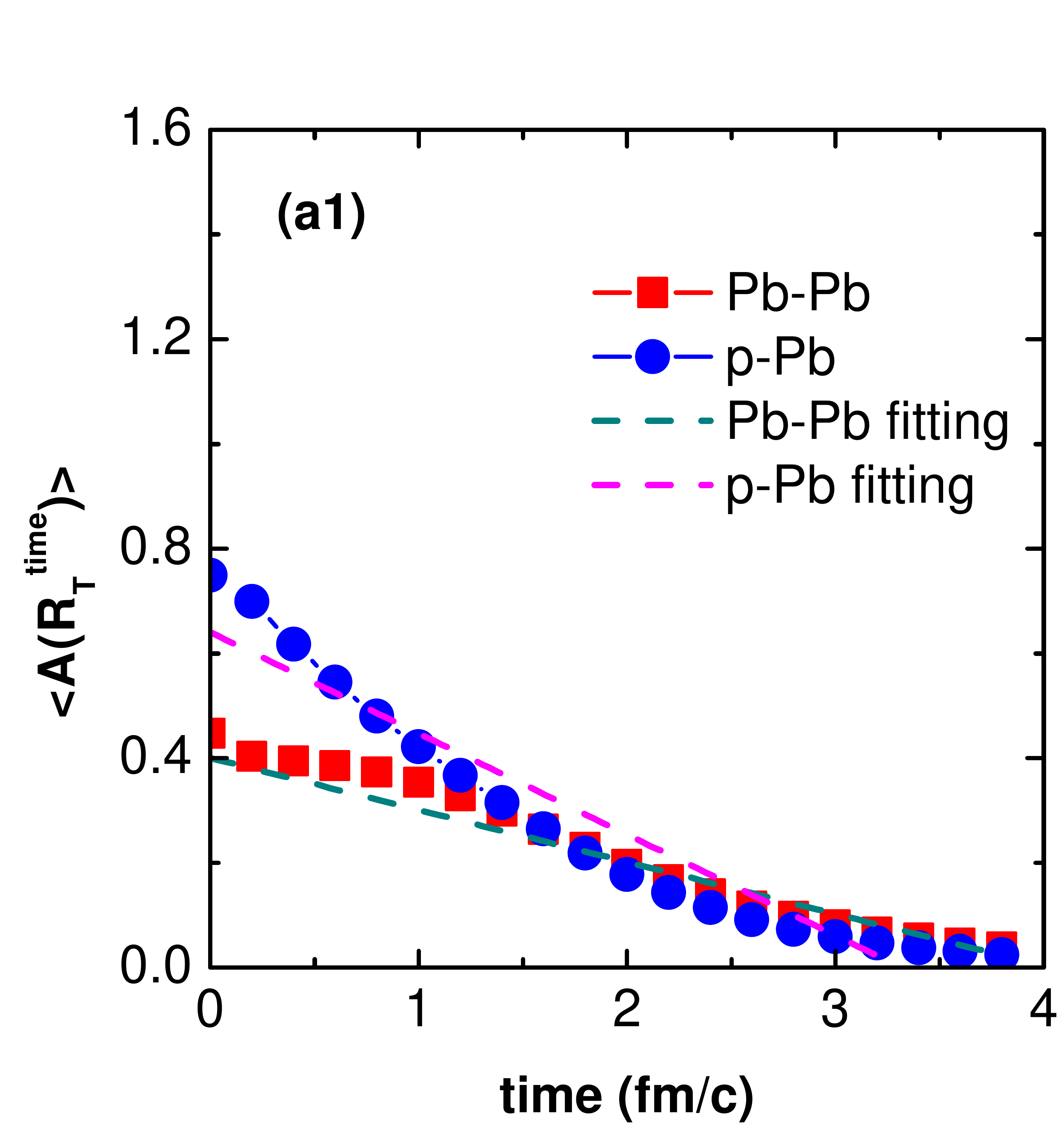}
\hspace{0.80cm}
\includegraphics[width=0.354\textwidth]{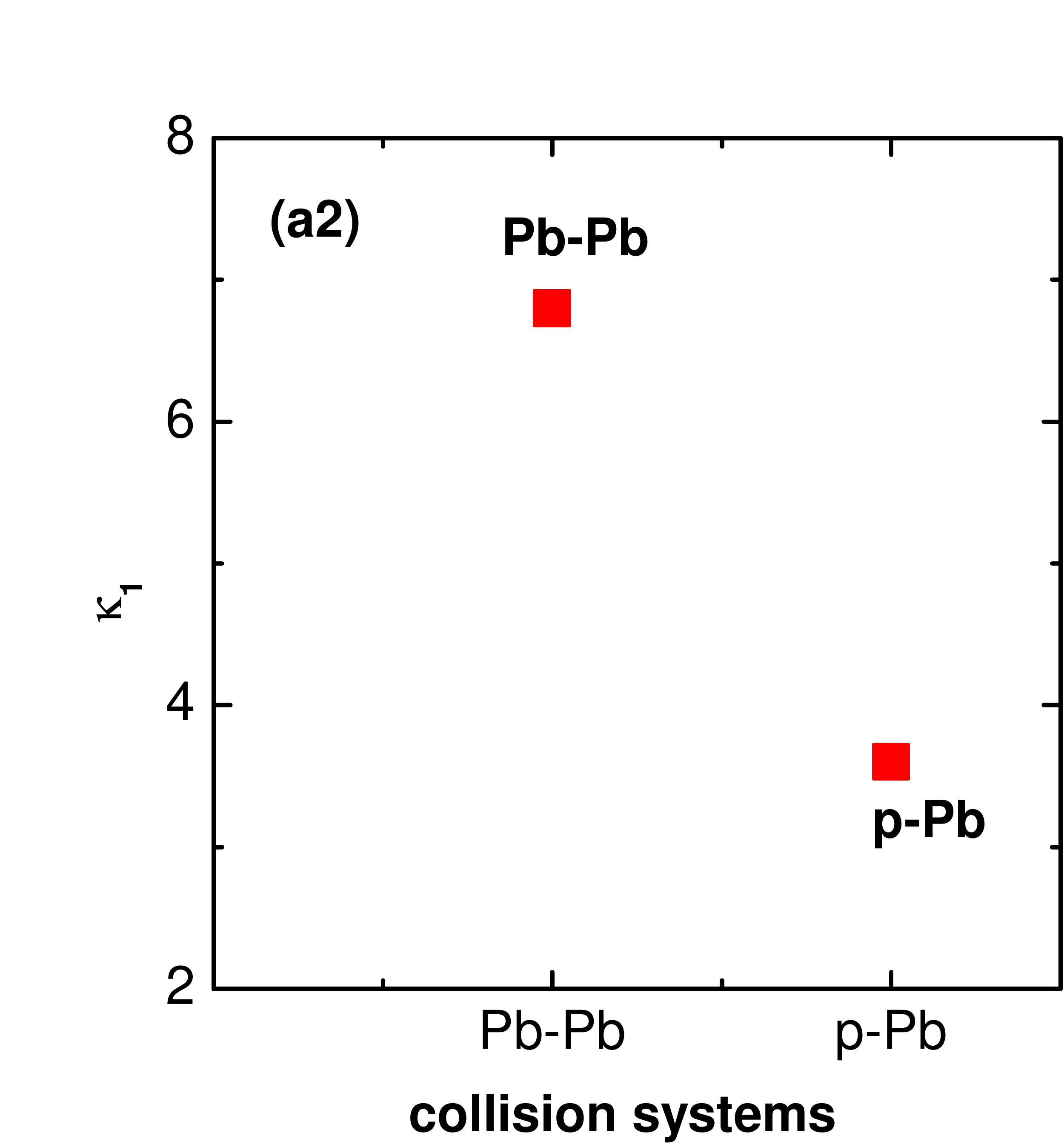}\\
\includegraphics[width=0.360\textwidth]{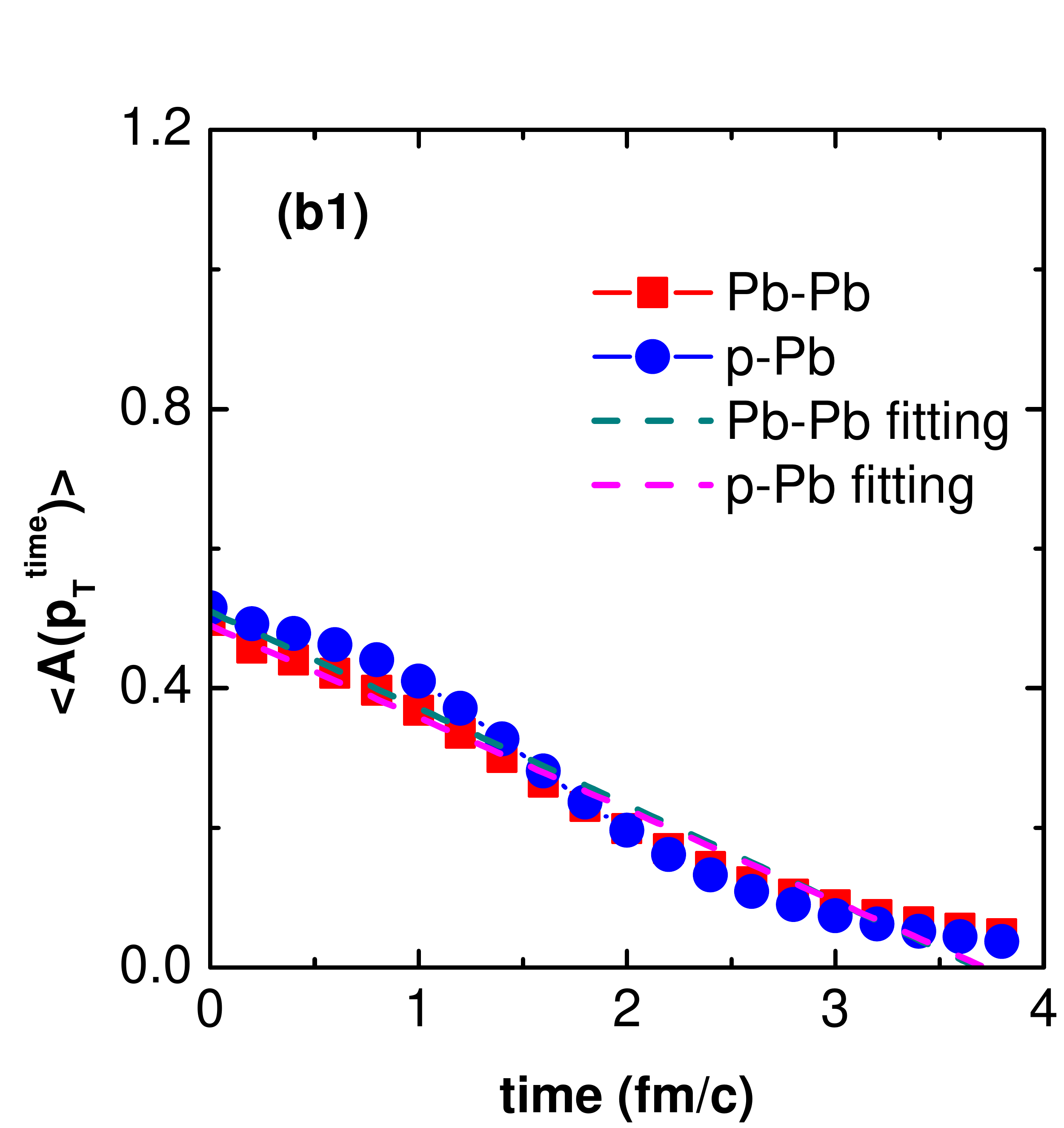}
\hspace{0.80cm}
\includegraphics[width=0.354\textwidth]{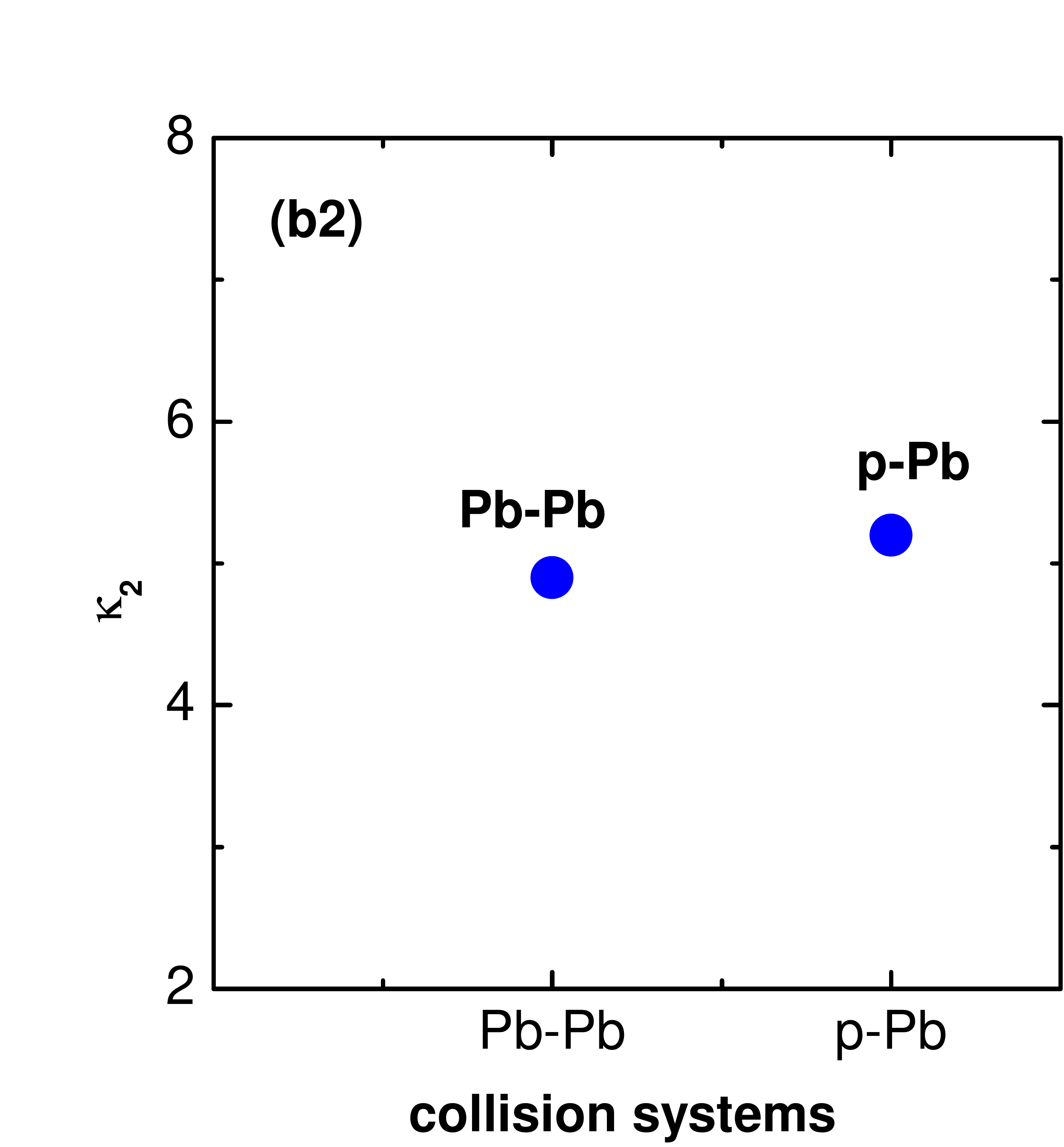}
\caption{(Color online)
Linear fitting of the averaged transverse shape asymmetry (up-left panel and down-left panel) as functions of the parton evolution time.
Linear response coefficients extract from \eq{resp:def1} and \eq{resp:def2} are shown in up-right panel and down-right panel, respectively.
Up panels: correlation between the final averaged transverse momentum asymmetry, $\langle A(p_{T}^{final})\rangle$, and time-dependent averaged transverse spatial asymmetry, $\langle A(R_{T}^{time})\rangle$.
Down panels: correlation between the final averaged transverse momentum asymmetry, $\langle A(p_{T}^{final})\rangle$, and time-dependent averaged transverse momentum asymmetry, $\langle A(p_{T}^{time})\rangle$.
}
\label{fig5}
\end{center}
\end{figure*}

\begin{figure*}[h]
\begin{center}
\includegraphics[width=0.480\textwidth]{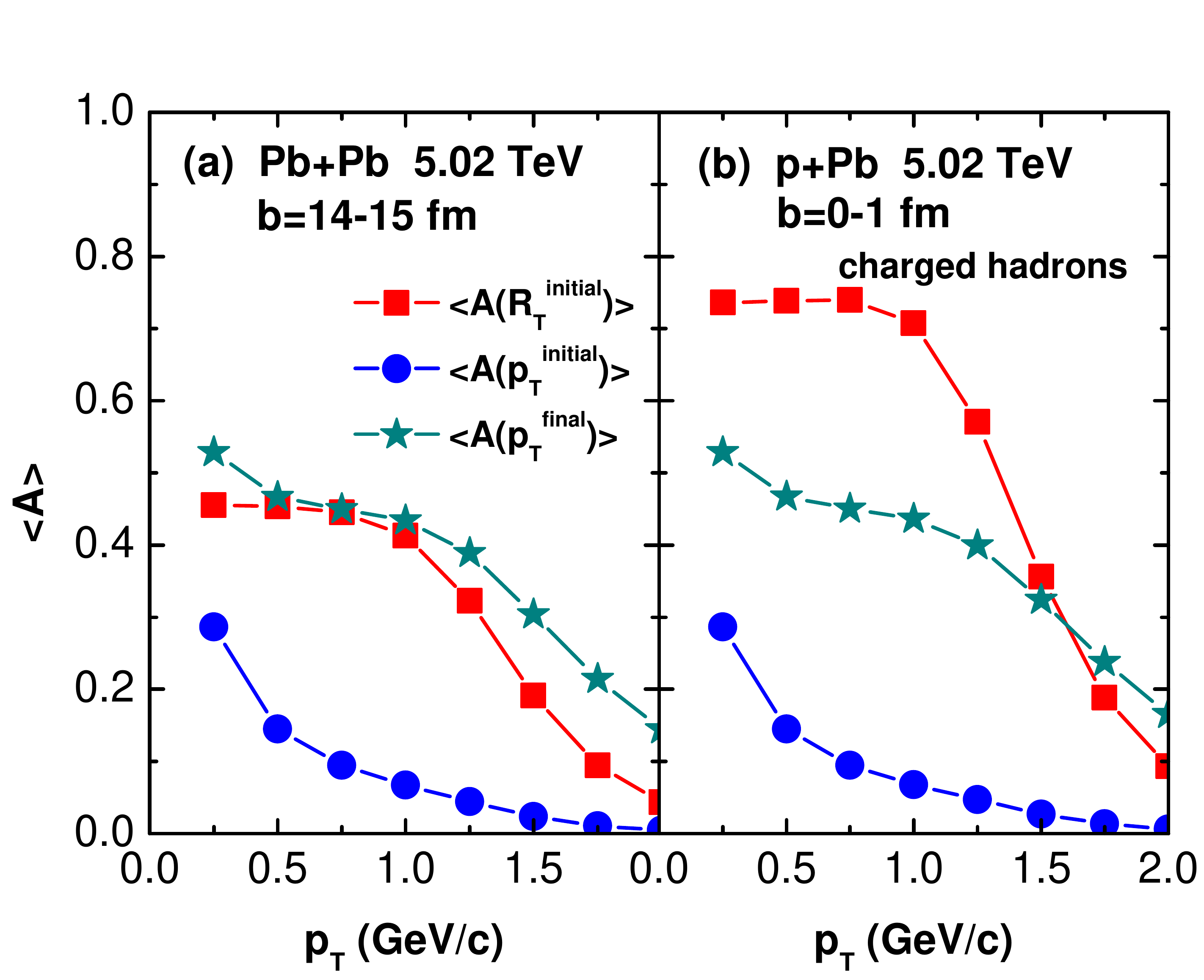}
\hspace{0.30cm}
\includegraphics[width=0.480\textwidth]{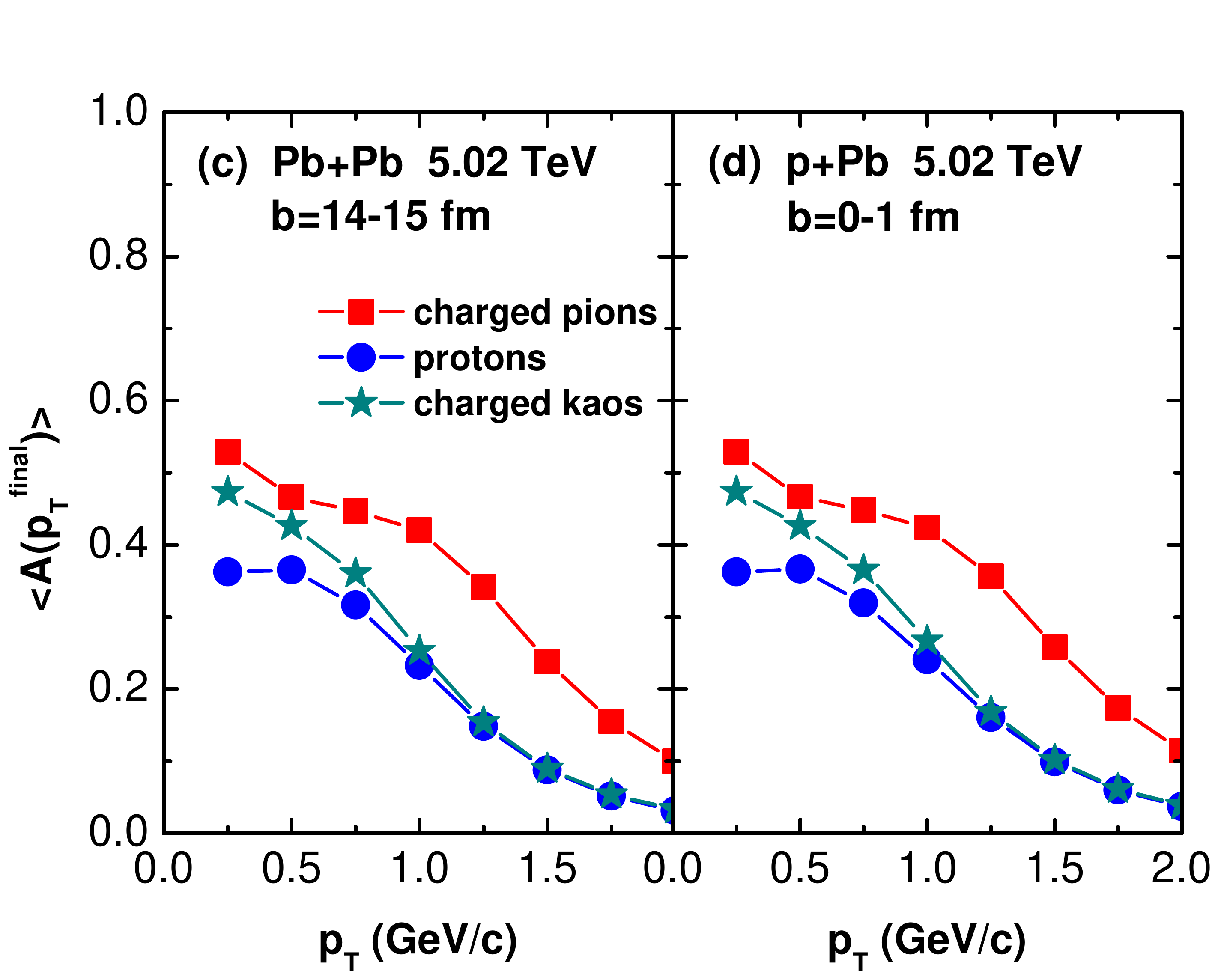}
\caption{(Color online)
Left panel: the averaged transverse asymmetry of initial spatial, initial momentum and final momentum, as functions of the transverse momentum $p_{T}$, where
(a) for Pb-Pb (b=14-15 fm) collisions and (b) for p-Pb (b=0-1 fm) collisions.
Right panel: the final averaged transverse momentum asymmetry are dependent on the PID in the present transverse momentum $p_{T}$ range, where
(c) for Pb-Pb (b=14-15 fm) collisions and (d) for p-Pb (b=0-1 fm) collisions.
}
\label{fig6}
\end{center}
\end{figure*}

The probability of parton multiplicity in small collisions probes the fluctuations of the initial state because they are to
a good approximation equal to the fluctuations of the initial total energy density of the system.
The results of probability and transverse asymmetry for Pb-Pb (b=14-15 fm) and p-Pb (b=0-1 fm) collisions at $\sqrt{s_{NN}}$= 5.02 TeV,
are shown in \Fig{fig1}.
The up-left panel of \Fig{fig1} shows the distribution of probability of initial parton multiplicity in Pb-Pb and p-Pb collisions,
and the up-right panel of \Fig{fig1} shows the distribution of probability of final charged hadron multiplicity in Pb-Pb and p-Pb collisions.
These initial parton multiplicity-dependent probability distributions are different in Pb-Pb and p-Pb collisions.
The peak of the distribution in Pb-Pb collisions is close to the zero parton multiplicity (similar to Ref.~\cite{Giacalone:2020eof}), while it is close to 200 in p-Pb collisions.
Such differences are similar to the final charged hadron multiplicity-dependent probability distributions,
i.e., the peak of the distribution in Pb-Pb collisions is close to multiplicity $=$ 10,
while it is close to 80 in p-Pb collisions, as shown in \Fig{fig1} (b).
We can conclude that the multiplicity distributions are different in those two systems.
Moving on to the spatial asymmetry, it also serves as a probe of the initial shape of the system,
a quantity that originates from the fluctuating geometry of the initial energy density fields.
In \Fig{fig1} (c) (down-left panel of \Fig{fig1}), the initial averaged transverse spatial asymmetry, $\langle A(R_{T}^{initial})\rangle$, as a function of the initial parton multiplicity, is shown.
The initial averaged transverse spatial asymmetry increases with the initial parton multiplicity increasing in p-Pb collisions,
while it is almost independent on the initial parton multiplicity in Pb-Pb collisions.
The value of $\langle A(R_{T}^{initial})\rangle$ is deviates significantly from zero in p-Pb collisions, and close to zero in Pb-Pb collisions.
These phenomena as a result of the degree of fluctuations of initial transverse, overlap in collisions.
Furthermore, the final averaged transverse momentum asymmetry, $\langle A(p_{T}^{final})\rangle$, as a function of the final charged hadron multiplicity,
is shown in \Fig{fig1} (d) (down-right panel of \Fig{fig1}).
Surprisingly, the value of $\langle A(p_{T}^{final})\rangle$ is almost consistent in Pb-Pb and p-Pb collisions.
These $\langle A(p_{T}^{final})\rangle$ depend weakly on the  final charged hadron multiplicity.
This reason can be explained by the hydrodynamics~\cite{Gardim:2021eoi}: even though the small initial fireball size is different,
but in the case of large initial entropy and energy,
it generates similarly large system in the final state.
To study this phenomenon, we will show the comparison of the initial,
time-dependent, and final shape asymmetry between Pb-Pb and p-Pb collisions.
We have chosen the same final charged hadron multiplicity in Pb-Pb and p-Pb collisions.
In this paper, for each chosen event, its final charged hadron multiplicity is ranged in $M_{ch}\in [60, 100]$.
Such settings are used in \cref{fig2,fig3,fig4,fig5,fig6}.

To estimate the asymmetry quantitatively, we show the initial and final shape asymmetry distributions for Pb-Pb and p-Pb systems in \Fig{fig2}.
In \Fig{fig2}, the up panels are the results of Pb-Pb (b=14-15 fm) collisions, and the down panels are the results of p-Pb (b=0-1 fm) collisions.
The x-y components of the initial spatial asymmetry are close to zero in Pb-Pb collisions [as in \Fig{fig2} (a1)],
while they deviates significantly from zero in p-Pb collisions [as in \Fig{fig2} (b1)].
It means that the initial spatial asymmetry distributions are dependent on the collision system.
However, the distributions of initial momentum asymmetry and final momentum asymmetry are similar both in Pb-Pb and p-Pb collisions,
as shown in \Fig{fig2} (a2), 2(b2), 2(a3) and 2(b3).
These effects can be understood as, in AMPT event-by-event simulations, the initial spatial asymmetry mainly comes from the initial geometric fluctuations of collisions,
and the initial momentum asymmetry mainly comes from the energy-momentum density fluctuations of collisions.

Hydrodynamics~\cite{Noronha-Hostler:2015dbi} showed that the final collective responds to the initial eccentricity.
And we can also see that both the initial transverse momentum asymmetry and final transverse momentum asymmetry are similar in Pb-Pb and p-Pb collisions.
In \Fig{fig3}, we show the correlation between the final transverse momentum asymmetry and initial transverse momentum asymmetry,
for Pb-Pb (b=14-15 fm) and p-Pb (b=0-1 fm) collisions at $\sqrt{s_{NN}}$= 5.02 TeV, respectively.
From \Fig{fig3}, there are weak correlations between the final transverse momentum asymmetry and initial transverse momentum asymmetry.
Such weak correlations are shown similarly in the two systems.
The linear correlation coefficients (noted Pearson coefficients) between the final transverse momentum asymmetry and initial transverse momentum asymmetry are $C[A(p_{T}^{initial}),~ A(p_{T}^{final})]$ (Pb-Pb) $\approx$0.21 and $C[A(p_{T}^{initial}),~A(p_{T}^{final})]$ (p-Pb) $\approx$0.18.
The magnitude of the Pearson coefficients is obtained by Ref.\cite{Wei:2020esd}.
These nonunity linear correlation coefficients also imply that the dynamic and/or dynamic fluctuations play an important role in the system evolution stage~\cite{Sakai:2021rdc}, i.e.,
the final transverse momentum asymmetry gets a large contribution from the initial fluctuations (noted by initial transverse momentum asymmetry)
and non-negligible contribution from the evolution of dynamics.

To study the contribution of dynamic and/or dynamic fluctuations,
the parton evolution time-dependent averaged transverse asymmetry are simulated. These results are shown in \Fig{fig4}.
The up panels of \Fig{fig4} are the results for Pb-Pb (b=14-15 fm) collisions at $\sqrt{s_{NN}}$= 5.02 TeV,
and the down panels of \Fig{fig4} are the results for p-Pb (b=0-1 fm) collisions at $\sqrt{s_{NN}}$= 5.02 TeV.
The x-y and transverse components of averaged spatial asymmetry and averaged momentum asymmetry are decrease with increasing evolution time.
In \Fig{fig4}, the x-y components of averaged asymmetry of spatial and momentum are similar in Pb-Pb and p-Pb collisions.
However, the ratios of the averaged transverse spatial asymmetry and averaged transverse momentum asymmetry are identifiable in the two systems.
Not only there is a weakly splitting of $\langle A(R_{T}^{time})\rangle$ and $\langle A(p_{T}^{time})\rangle$ in Pb-Pb collisions, but also there is a remarkable splitting in the p-Pb collisions.
The reason for the splitting phenomenon is that orderly heating of a nonequilibrium system is required during the initial short time to boost-invariant gluon fields expansion.
While the time is approximately equal to 1.0 fm/c, the system reaches the state of local thermal equilibrium and
enters the hydrodynamic stage with macroscopic properties of temperature~\cite{Heinz:2002eta,Schlichting:2018etd,Gale:2020Ped,Bhalerao:2015cfi}.
The system in the thermal equilibrium state has a symmetrical transverse expansion rate, as a result, it has a similarly transverse asymmetry rate,
i.e., $\langle A(p_{T}^{time})\rangle$ (Pb-Pb) $\approx \langle A(p_{T}^{time})\rangle$ (p-Pb) with $\langle A(R_{y})\rangle/\langle A(R_{x})\rangle \approx 1$.
These results are shown in \Fig{fig4} (a3) and (b3).

From \Fig{fig3}, the final transverse momentum asymmetry is independent of the initial transverse momentum asymmetry, both in the Pb-Pb and p-Pb systems.
The time-dependent transverse shape asymmetry are shown smoothly in \Fig{fig4}.
A natural idea is that the final transverse momentum asymmetry depend on the gradient of time.
If $\langle A(R_{T}^{time})\rangle$ begins in \Fig{fig4} and $\langle A(p_{T}^{final})\rangle$, we can get a linear response from these two quantities, as
\begin{eqnarray}\label{resp:def1}
\langle A(p_{T}^{final})\rangle &\approx& -\kappa_{1}\frac{\partial \langle A(R_{T}^{time})\rangle}{\partial t}.
\end{eqnarray}

Similarly, if $\langle A(p_{T}^{time})\rangle$ begin in \Fig{fig4} and $\langle A(p_{T}^{final})\rangle$, we can also get a linear response from these two quantities, as
\begin{eqnarray}\label{resp:def2}
\langle A(p_{T}^{final})\rangle &\approx& -\kappa_{2}\frac{\partial \langle A(p_{T}^{time})\rangle}{\partial t}.
\end{eqnarray}

In this work, we only study the degree of linear correlation of the samples, without entering the discussion about the possible nonlinear correlations.
Given $\langle A(p_{T}^{final})\rangle$~(Pb-Pb)$\approx$0.67 and $\langle A(p_{T}^{final})\rangle$~(p-Pb)$\approx$0.69 from the final charged hadrons,
we can extract the response coefficient $\kappa_{1}$  and $\kappa_{2}$ from \eq{resp:def1} and \eq{resp:def2}, respectively.
The linear fitting of the averaged transverse asymmetry, both of the spatial and momentum as functions of the parton evolution time,
is shown in \Fig{fig5} (a1) and (b1). The linear response coefficients extracted from \Fig{fig5} (a1) and (b1)
are significantly dependent on the collision system, as shown in \Fig{fig5} (a2) and (b2), respectively.
From \Fig{fig5} (a2) and (b2), one can see that the response coefficients are significantly deviate from zero both in Pb-Pb and p-Pb collisions.
These nonzero linear response coefficients imply that the final momentum asymmetry is significantly dependent on the evolution of parton dynamics.
The linear response coefficient $\kappa_{1}$ is identifiable in Pb-Pb and p-Pb collisions, while $\kappa_{2}$ is almost identical in those two systems.

To provide effective observability to the experiment, we show the transverse momentum and the particle species identity (PID)-dependent averaged transverse shape asymmetry in this work.
In \Fig{fig6}, the averaged transverse shape asymmetry is significantly dependent on the transverse momentum (a) and PID (b) in Pb-Pb and p-Pb collisions.
For each $p_{T}$ bin, there is ranged in the same final charged hadron multiplicity $M_{ch}\in [60, 100]$.
From \Fig{fig6} (a) and (b) (left panel), both the initial and final averaged transverse shape asymmetry decrease with increasing transverse momentum ($p_{T}$).
Such phenomena are similar in Pb-Pb and p-Pb collisions, while the peak of the initial averaged transverse spatial asymmetry is different.
Again, the initial spatial asymmetry mainly comes from the initial geometric fluctuations of the event-by-event collisions,
and the initial momentum asymmetry mainly comes from the energy-momentum density fluctuations of collision.
From \Fig{fig6} (c) and (d) (right panel), the final averaged transverse momentum asymmetry is significantly dependent on the PID,
i.e., the scales ordering of final averaged transverse momentum asymmetry is $\langle A(p_{T}^{final})\rangle$(charged pions)$\geq \langle A(p_{T}^{final})\rangle$(charged kaos)$\geq \langle A(p_{T}^{final})\rangle$(protons). These final averaged transverse momentum asymmetry of all PID are also decrease with increasing the transverse momentum ($p_{T}$).
For lower $p_{T}$ with larger $\langle A(p_{T}^{final})\rangle$, the reason is that the lower $p_{T}$ shape undergoes, the more about the soft scattering process in the evolution,
as a result, it generates more dynamic fluctuations. Such $p_{T}$ dependent and PID scales ordering phenomena is similarly in Pb-Pb and p-Pb collisions.
For the same multiplicity, the final averaged transverse momentum asymmetry are PID scales ordering means that the protons are influenced by the density/volume fluctuations in collisions.
It can be similarly understood that the final collective coefficient $v_{2}$ are scales ordering on PID~\cite{Abelev:2013lac} due to the initial density/volume fluctuations.

\section{Summary}
\label{sec:sum}
In summary, based on AMPT event-by-event calculations, this paper carried out the parton evolution time-dependent transverse asymmetry
in Pb-Pb (b=14-15 fm) and p-Pb (b=0-1 fm) collisions at $\sqrt{s_{NN}}$= 5.02 TeV, respectively.
The results showed that the initial spatial asymmetry is dependent on the collision system,
while both the initial and final momentum asymmetry are independent on the collision system.
By the evolution time step, we extracted the linear response coefficients from quantities between the final momentum asymmetry and time-dependent momentum asymmetry.
The nonzero extraction response coefficients imply that the final momentum asymmetry significantly depends on the evolution of parton dynamics.
That is the reason the final momentum asymmetry is weakly correlated with initial momentum asymmetry.
Furthermore, the averaged transverse shape asymmetry, which includes the initial spatial asymmetry,
the initial and final momentum asymmetry, is significantly dependent on the transverse momentum and PID.
The PID scales ordering transverse asymmetry indicated that it provides a possible observation
for studying the fluctuating properties of QGP droplets, which may produce in heavy-ion collisions.
Such possible QGP droplets are fascinating, but they still are open questions and opportunities for future improvements.

\section*{Acknowledgements}
This work are supported by the Youth Program of Natural Science Foundation of Guangxi (China), with Grant No.~2019GXNSFBA245080, the Special fund for talentes of Guangxi (China), with Grant No.~AD19245157, and also by the Doctor Startup Foundation of Guangxi University of Science and Technology, with Grant No.~19Z19.


\end{document}